\documentclass[
floatfix, 
pra, 
twocolumn, 
reprint, 
amsmath,amssymb, 
aps, 
footnote, 
]{revtex4}
\usepackage{graphicx}
\usepackage{dcolumn}
\usepackage{bm}
\usepackage{multirow}
\usepackage{color}
\usepackage{braket}
\usepackage[utf8]{inputenc}

\newcommand{\sig}{\sigma}

\newcommand{\mr}{\mathrm}

\newcommand{\pri}{^\prime}

\newcommand{\mbZ}{\mathbb{Z}}

\newcommand{\mbT}{\mathbb{T}}

\newcommand{\nnco}{\nonumber \\}

\newcommand{\peqn}[1]{
\begin{align}
#1
.\end{align}
}
\newcommand{\ceqn}[1]{
\begin{align}
#1
,\end{align}
}
\newcommand{\neqn}[1]{
\begin{align}
#1
\end{align}
}

\newcommand{\tbr}[1]{
\left\{ #1 \right\}
}
\newcommand{\rbr}[1]{
\left[ #1 \right]
}
\newcommand{\br}[1]{
\left( #1 \right)
}

\newcommand{\twmat}[4]{
\left(
\begin{array}{cc}
#1 & #2 \\
#3 & #4
\end{array}
\right)
}

\newcommand{\bk}{\bm{k}}
\newcommand{\Exp}[1]{\mr{e}^{#1}}
\newcommand{\eps}{\epsilon}
\newcommand{\heff}{h_{\mr{eff}}}

\newcommand{\wt}{\widetilde}

\begin{document}

\preprint{APS/123-QED}
\title{Floquet chiral magnetic effect} 
\author{Sho Higashikawa${}^1$}
\author{Masaya Nakagawa${}^1$}
\author{Masahito Ueda${}^{1,2}$}
\affiliation{
${}^1$
Department of Physics, University of Tokyo, 7-3-1 Hongo, Bunkyo-ku, Tokyo 113-0033, Japan \\
${}^2$RIKEN Center for Emergent Matter Science (CEMS), Wako, Saitama 351-0198, Japan
}
\date{\today}

\begin{abstract}
A single Weyl fermion, which is prohibited in static lattice systems by the Nielsen-Ninomiya theorem, is shown to be realized in a periodically driven three-dimensional lattice system with a topologically nontrivial Floquet unitary operator, manifesting the chiral magnetic effect. 
We give a topological classification of Floquet unitary operators in the Altland-Zirnbauer symmetry classes for all dimensions, 
and use it to predict that all gapless surface states of topological insulators and superconductors can emerge in bulk quasienergy spectra in Floquet systems.
\end{abstract}
\pacs{Valid PACS appear here}
\maketitle


In 1981, Nielsen and Ninomiya proved that a single Weyl fermion cannot be realized in lattice systems \cite{NielsenNinomiya1, NielsenNinomiya2}. This theorem places a fundamental constraint on band structures due to the topology of the Brillouin zone. 
Weyl fermions have recently played a key role in cross-fertilizing ideas from high-energy physics and condensed-matter physics. 
A prime example is the prediction of Weyl semimetals \cite{Murakami07, Wan11}, where the low-energy effective field theory of Weyl fermions predicts novel electromagnetic responses originating from the chiral anomaly \cite{NielsenNinomiya3, Zyuzin12, Son12, Burkov14}. 
In particular, the observations of the surface Fermi arc \cite{Lv15_1, SYXu15, Lv15_2, LXYang15, SYXu16, NXu16, SMHuang15} and anomalous transport \cite{XHuang15, Arnold16, Zhen16, ZhangHasan16} have aroused considerable interest. 
However, if a system is defined on a lattice and thus anomaly-free, the Nielsen-Ninomiya theorem dictates that a Weyl fermion be accompanied by its partner with opposite chirality. 
By the same token, an anomaly-induced response known as the chiral magnetic effect (CME) \cite{Fukushima08} does not occur in equilibrium \cite{Vazifeh13}, 
and numerous attempts to circumvent this difficulty have been made \cite{Goswami15, Taguchi16, Sumiyoshi16, Ibe17, Cortijo16, Pikulin16, MengT18, ChernodubM19}.

In this Letter, we demonstrate that a single Weyl fermion can be realized on a periodically driven lattice, thereby overcoming the above limitations. 
In periodically driven (Floquet) systems, the unitary time-evolution operator over one period defines an effective Hamiltonian and the associated quasienergies \cite{Eckardt17}. 
Despite the apparent similarity to static systems, Floquet systems enable the realization of exotic phases that cannot be achieved in equilibrium, such as anomalous  topological insulators \cite{Kitagawa10, Rudner13} and time crystals \cite{Else16, Yao17}. 
The key idea of our proposal is an emerging topological structure in unitary operators associated with the periodicity of quasienergies \cite{Kitagawa10}. 
We here show that a driving protocol for a three-dimensional (3D) Thouless pump \cite{Thouless83, Kitagawa10} given by a topological Floquet unitary operation realizes a single Weyl fermion in a 3D lattice system, thereby providing a platform to observe the CME. 
We demonstrate that chiral transport emerges under a topological 3D Floquet drive with an applied synthetic magnetic field, leading to a Floquet realization of the CME. 
Our proposal can be implemented by using ultracold atomic gases, where the Thouless pump has been realized experimentally \cite{Nakajima16, Lohse16}. 

Furthermore, by exploiting the correspondence between anomalous gapless spectra and topological unitary operators, we provide a topological classification of Floquet unitary operators in the Altland-Zirnbauer symmetry classes. 
In general, a wide variety of lattice-prohibited band structures under given symmetries can be realized as gapless surface states of topological phases \cite{QiZhang11}. 
The impossibility of pure lattice realization of surface states is deeply connected with their symmetry-protected gaplessness via quantum anomalies \cite{RyuZhang12, Sule13, Hsieh16}. 
We show that the classification of topological Floquet unitaries, which generically offer symmetry-protected gapless quasienergy spectra, coincides with that of gapless surface states of static topological insulators/superconductors (TIs/TSCs). This correspondence strongly suggests that one can realize any gapless surface states of TIs/TSCs as \textit{bulk} quasienergy bands of Floquet systems, even though they cannot be realized in static lattice systems. 

\textit{General strategy.}--\ 
We first explain our strategy to obtain a lattice-prohibited band structure in a Floquet system. We consider a periodically driven system of non-interacting fermions on a lattice. A periodically driven system is characterized by a Floquet operator, which is defined as a time-evolution operator over one period \cite{Eckardt17}. Since a crystal momentum is a good quantum number, a Floquet operator defines a map from the Brillouin zone to a space of unitary matrices. To characterize topology of a Floquet operator, we assume that a Hilbert subspace of the system is mapped onto itself by the Floquet operator \cite{Kitagawa10}. This condition is achieved by, e.g., (i) using generalized adiabaticity \cite{Kitagawa10, SunX18}, in which a time evolution is restricted to a low-energy subspace due to a large separation between low and high energy bands, or (ii) some fine-tuning of a driving protocol \cite{Budich17}. We hereafter refer to the Hilbert subspace closed in the time evolution over one period as ``lower" Floquet bands, which play a role similar to occupied bands of static  insulators \cite{Kitagawa10}. 

Let us denote the Floquet operator restricted to the lower Floquet bands by $U(\bm{k})$, where $\bm{k}$ is a crystal momentum. When $U(\bm{k})$ offers a topologically nontrivial map from the Brillouin zone to a unitary group U($N$) ($N$ is the number of the lower Floquet bands), the lower Floquet bands possess gapless quasienergy spectra, since a gapped Floquet operator can continuously be deformed into a trivial unitary, e.g., $U(\bm{k})=\bm{1}_N$ \cite{Kitagawa10, NathanRudner15}. Since the gapless quasienergy spectra cannot be gapped out by a continuous deformation of the Floquet operator, a topological Floquet operator is expected to exhibit a topologically protected gapless band structure, such as the case of Weyl fermions. 
In fact, it has been shown \cite{Kitagawa10, 1d_winding} that a single chiral fermion, which is forbidden in a static one-dimensional lattice, can be realized with a Floquet operator that has a nontrivial winding number. 
Our strategy for a realization of a single Weyl fermion is to construct a driving protocol that gives a Floquet operator with a nontrivial topological number in a 3D lattice system.

\begin{figure}[t]
  \begin{center}
	\includegraphics[clip, width = \columnwidth]{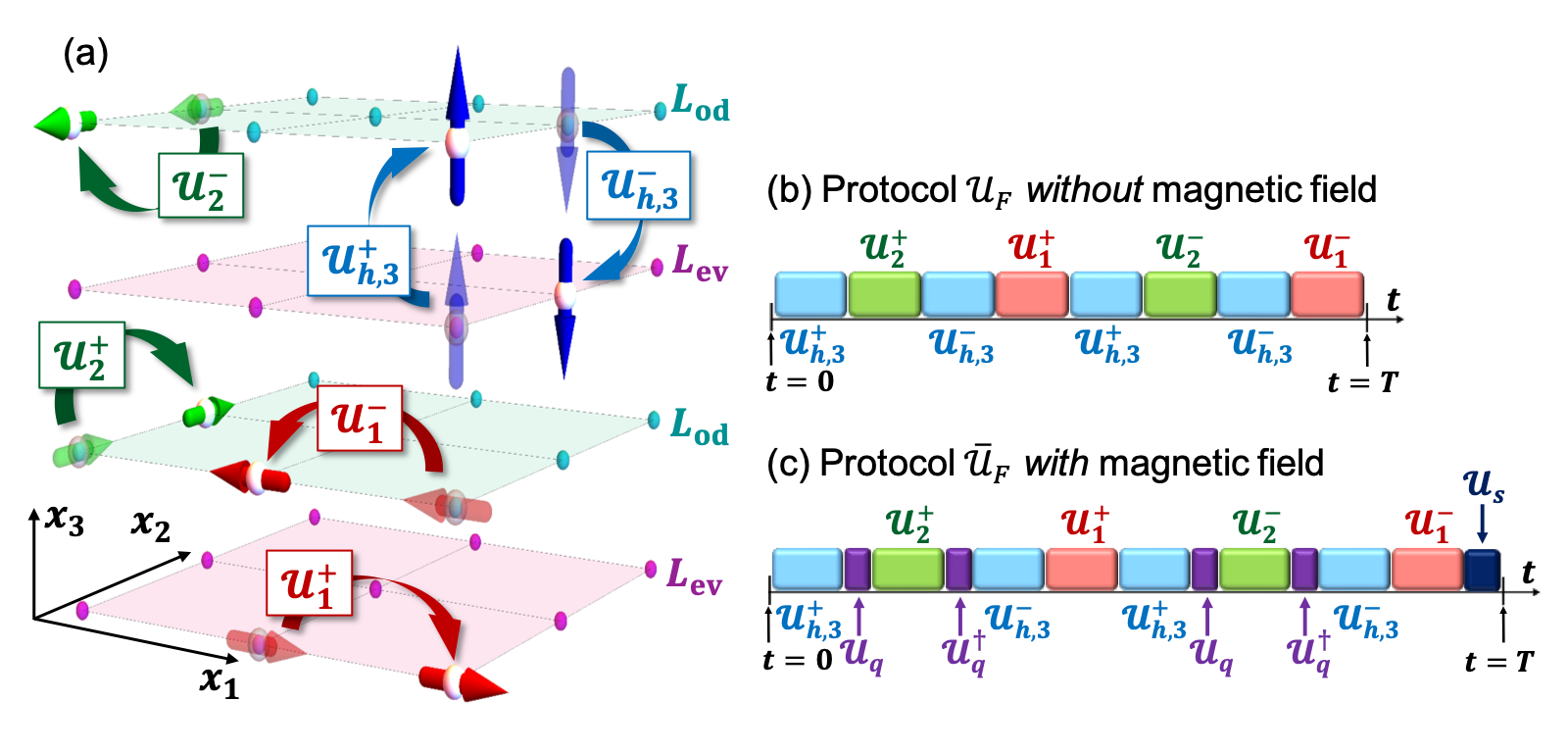}
	\caption{
(a) Schematic illustration of the fermion pump by time-evolution operators $\mathcal{U}_j^\pm \ (j=1,2)$ and $\mathcal{U}_{h,3}^\pm$ given in Eqs.~\eqref{eq: in-plane pump} and \eqref{eq: in-axis pump} in the 3D lattice with a sublattice structure $L_{\mr{ev}}$ and $L_{\mr{od}}$ in the third direction. Thick arrows show the spin directions of fermions. 
(b), (c) Driving protocols of the model (b) without and (c) with a magnetic field. 
The time-evolution operator $\mathcal{U}_q$ describes a sudden switch-on and -off of a quadrupole potential, and $\mathcal{U}_s$ describes hopping in the third direction with a Hamiltonian $H_s$.}
	\label{fig: ModelAndProtocol}
  \end{center}
\end{figure}

\textit{Model.}--\ 
Let us now proceed to a construction of a model with a topological Floquet operator. 
We consider spin-half fermions on a cubic lattice $L_C$ with a sublattice structure in the third direction: 
\begin{align}
	L_C := \tbr{\left. \br{m_1, m_2, {m_3 \over 2}}\right| m_1, m_2, m_3 \in \mbZ }
	,
\end{align}
where the sublattice $L_{\mr{ev}}$ ($L_{\mr{od}}$) corresponds to the sites with even (odd) $m_3$ [purple (light blue) points in Fig.~\ref{fig: ModelAndProtocol} (a)].
The lattice constant $a_{\mr{lat}}$ is set to be unity: $a_{\mr{lat}} = 1$. 
The main ingredient of our model is spin-selective Thouless pumps \cite{Kitagawa10, Thouless83, Budich17} whose time-evolution operators $\mathcal{U}_j^\pm \ (j= 1,2)$ are given by 
\begin{align}
	\mathcal{U}_j^\pm :=& \sum_{\bm{x} \alpha,\beta} 
\rbr{
(P_j^\pm)^{\alpha\beta} c_{\bm{x} \pm \bm{e}_j, \alpha}^\dagger c_{\bm{x}, \beta}
 + (P_j^\mp)^{\alpha\beta} c_{\bm{x}, \alpha}^\dagger c_{\bm{x}, \beta}
}
	, \label{eq: in-plane pump}
\end{align} 
where $\bm{x} = (x_1, x_2, x_3) \in L_C$ denote the lattice site, $\bm{e}_j$ is a unit vector in the $x_j$ direction, and
$c_{\bm{x}} = (c_{\bm{x}, \uparrow}, c_{\bm{x}, \downarrow})$ is the annihilation operator of a fermion with spin $\alpha$ ($\uparrow$ or $\downarrow$) at site $\bm{x}$. 
The matrix $P_j^\pm := (\sig_0 \pm \sig_j)/2$ is a projection operator on a spin state $\sig_j = \pm 1$, with $\sig_0$ and $\sig_j \ (j=1,2,3)$ being the $2 \times 2$ identity matrix and the Pauli matrices. 
From the projective nature of $P_j^\pm$, under the pump $\mathcal{U}_j^+$ ($\mathcal{U}_j^-$), 
fermions in a spin state $\sig_j = +1$ ($-1$) are displaced by one lattice site in the positive (negative) $x_j$ direction, while fermions in a spin state $\sig_1 = -1$ ($+1$) are not, thereby achieving spin-selective transport [see red and green arrows in Fig.~\ref{fig: ModelAndProtocol} (a)]. 
We also introduce spin-selective Thouless pumps $\mathcal{U}_{h,3}^\pm$ which displace fermions by a half lattice site in the $x_3$ direction: 
\begin{align}
	\mathcal{U}_{h,3}^\pm &:= 
	\sum_{\bm{x}, \alpha,\beta}	
	\rbr{
(P_j^\pm)^{\alpha\beta} c_{\bm{x} \pm {\bm{e}_3 \over 2}, \alpha}^\dagger c_{\bm{x}, \beta}
 + (P_j^\mp)^{\alpha\beta} c_{\bm{x}, \alpha}^\dagger c_{\bm{x}, \beta}
	}
	. \label{eq: in-axis pump}
\end{align}
We note that fermions can be displaced between the unit cells by $\mathcal{U}_{h,3}^\pm$ [see blue arrows in Fig.~\ref{fig: ModelAndProtocol} (a)]. 

The driving protocol of our topological pump is constituted from eight successive applications of $\mathcal{U}_1^\pm, \mathcal{U}_2^\pm$, and $\mathcal{U}_{h, 3}^\pm$ as shown in Fig.~\ref{fig: ModelAndProtocol} (b), where the total time-evolution operator $\mathcal{U}_F^{\mr{wh}}$ for the whole four bands over one cycle is given as follows \cite{3d_winding, experiment}: 
\begin{align}
	\mathcal{U}_F^{\mr{wh}} &:= 
\mathcal{U}_1^- \mathcal{U}_{h,3}^- \mathcal{U}_2^- \mathcal{U}_{h,3}^+ 
\mathcal{U}_1^+ \mathcal{U}_{h,3}^- \mathcal{U}_2^+ \mathcal{U}_{h,3}^+ = 
	\sum_{\bk} c_{\bk}^\dagger V^{\mr{wh}}(\bk) c_{\bk}
	, \label{eq: whole protocol}
	\end{align}
where $\bk = (k_1, k_2, k_3)$ is the crystal momentum. 
Then, the Floquet operator $V^{\mr{wh}}(\bk)$ is decomposed into two $2 \times 2$ matrices: $V^{\mr{wh}}(\bk) = U(\bk) \oplus U^H(\bk)$, where
\begin{align}
	U(\bk) &:= 
	U_1^-\br{k_1}U_{h,3}^-\br{k_3}U_2^-\br{k_2}U_{h,3}^+\br{k_3}
	\nonumber \\ & \quad\quad \times 
	U_1^+\br{k_1}U_{h,3}^-\br{k_3}U_2^+\br{k_2}U_{h,3}^+\br{k_3}
	, \label{eq: strobo0} 
\end{align}
and $U^H(\bk):= U(k_1, k_2, k_3-2\pi)$, $U_{j}^\pm(k) := P_{j}^\pm \mr{e}^{-ik} + P_{j}^\mp$ and $U_{h,3}(k) := U_3(k/2)$ represent the Floquet operators of the spin-selective Thouless pumps. 
Here we focus on $U(\bm{k})$ as a Floquet operator of lower Floquet bands. 
A straightforward calculation shows that $U(\bk)$ stays a constant value $-\sig_0$ if $\bk$ belongs to the boundary of the Brillouin zone $\mbT^3 := [-\pi,\pi]^3$ and hence satisfies the periodic boundary condition on $\mbT^3$. 

Let $\heff(\bk)$ be an effective Hamiltonian defined by $U(\bk) =: \exp\rbr{ -i \heff (\bk)}$, where the driving period $T$ is set to unity. 
Since the quasienergies $\epsilon(\bk)$ (eigenvalues of $\heff(\bk)$) satisfy $\cos \rbr{\epsilon(\bk)} = \mr{Tr}\rbr{U(\bk)} / 2$, 
$\epsilon(\bk) = 
\pm\cos^{-1}\rbr{
2 \cos^2\br{{k_1 \over 2}} \cos^2\br{{k_2 \over 2}} \cos^2\br{{k_3 \over 2}} - 1
}$ follows from Eq.~\eqref{eq: strobo0}. 
Therefore, $\epsilon(\bk)$ has only one gapless point at $\bk = 0$. 
Since $U(\bk)$ is expanded around $\bk=0$ as $U(\bk) \approx \sig_0 - i \bk \cdot \bm{\sig}$ from $U_j^\pm(k) \approx \sig_0 \mp i P_j^\pm k$ for $k \approx 0$, we have $\heff(\bk) \approx \bk \cdot \bm{\sig},$ which clearly indicates the presence of a single left-handed Weyl fermion. 
The presence and stability of this single Weyl fermion is protected by a nontrivial topology in $U(\bk)$; In fact, $U(\bk)$ achieves a topologically nontrivial map from $\mbT^3$ to $\mr{SU}(2) \cong S^3$ (3D sphere) with a unit winding number \cite{MantonN04}:
\begin{align}
	W := - \int {d\bk \over 24 \pi^2} 
	\sum_{i,j,k = 1}^3
	\epsilon^{ijk} \mr{Tr} \left[ R_i R_j R_k \right]
	 = 1
	, \label{eq: windnumber}
\end{align}
where $R_i := U(\bk)^\dagger \partial_{k_i} U(\bk)$ \cite{3d_winding}. 
We note that our model should be distinguished in topology from the previous proposals for realizing Floquet Weyl semimetals \cite{Chan16a, Hubener17, WangH16, Zou16, Bucciantini17, Wang14, Ebihara16, ZhangNagaosa16, Yan16, Takasan17}, where the Weyl nodes always appear in pairs in accordance with the Nielsen-Ninomiya theorem.

\begin{figure}[t]
  \begin{center}
	\includegraphics[clip, width = \columnwidth]{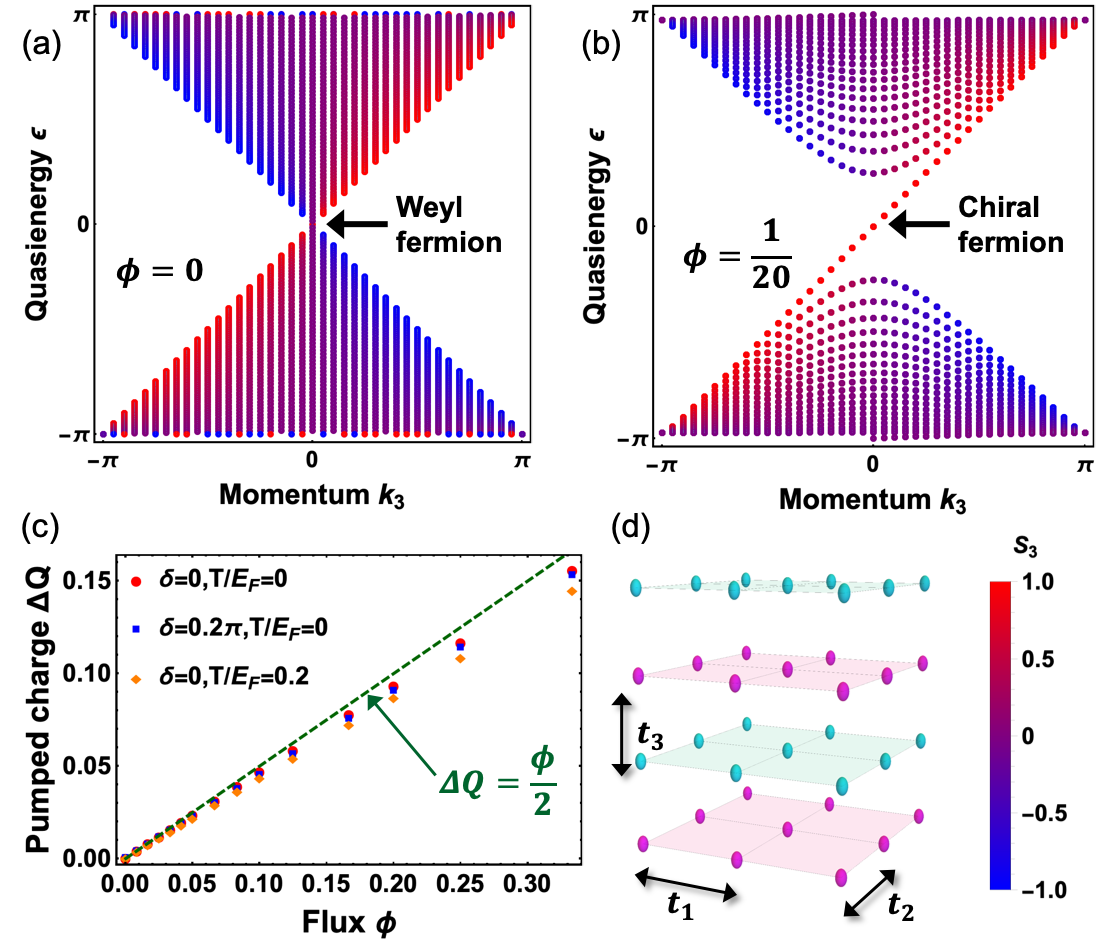}
	\caption{
(a), (b) Quasienergy spectra of $\bar{U}(0, k_3)$ with flux (a) $\phi=0$ and (b) $\phi=1/20$. The color represents the expectation value $S_3$ of $\sig_3$ for each eigenstate according to the gauge shown on the right. 
(c) Pumped charge $\Delta Q$ as a function of flux $\phi$ for the initial state \eqref{eq: fcme, 8} at half-filling and zero temperature without (red) and with (blue) spin-mixing perturbation. The orange points are the results with finite-temperature $T = 0.1 E_F$ [$E_F$ is the Fermi energy]. 
(d) Hopping amplitudes $t_i$ for preparing the initial state.}	\label{fig: DynaWmag1}
  \end{center}
\end{figure}

\textit{Floquet chiral magnetic effect.}--\ 
When a magnetic field is applied, a Weyl fermion shows chiral transport antiparallel to the applied magnetic field, a phenomenon known as the CME \cite{Fukushima08}. 
A magnetic field can be introduced in our model through the replacement of 
$\mathcal{U}_2^\pm$ in Eq.~\eqref{eq: strobo0} with $\mathcal{U}_q^\dagger \mathcal{U}_2^\pm \mathcal{U}_q$, where $\mathcal{U}_q := \exp(- i 2 \pi \phi x_1 x_2)$ is the time evolution operator induced by a sudden switch-on and -off of a quadrupole potential \cite{Sorensen05, HafeziM07}. 
Since $\mathcal{U}_q^\dagger U_2^\pm(k_2) \mathcal{U}_q = U_2^\pm(k_2 - 2 \pi \phi x_1)$, the effective Hamiltonian near $\bm{k} = 0$ is given by $h_{\mr{eff}} = (\bk + \bm{A})\cdot \bm{\sigma}$ with $\bm{A} = (0, -2 \pi \phi x_1, 0)$, which describes a Weyl fermion under a magnetic field $\bm{B} = (0,0, -2 \pi \phi)$. 
However, $\mathcal{U}_q$ couples the lower and higher Floquet bands. 
Therefore, to decouple them, we need an additional time evolution $\mathcal{U}_s$ with duration $\tau_s$ under a Hamiltonian $H_s = J_s \sum_{\bm{x}} (i c_{\bm{x} + {\bm{e}_3 \over 2}}c_{\bm{x}} + \mr{h.c.})$ with $4J_s\tau_s = \pi\phi$ at the end of the cycle [see Fig.~\ref{fig: ModelAndProtocol} (c)] \cite{experiment}. 
Since $k_2$ and $k_3$ remain as good quantum numbers, the Floquet operator $\bar{U}_F$ acting on the lower Floquet band is decomposed into a set of one-dimensional lattice models as $\bar{U}_F = \sum_{k_2, k_3} \bar{U}(k_2, k_3)$ \cite{barU_form}. 
Figures~\ref{fig: DynaWmag1} (a) and (b) show the quasienergy spectra of $\bar{U}(0, k_3)$ without ($\phi = 0$) and with ($\phi = 1/20$) a magnetic field, respectively, where the color of the points represents the spin polarization $S_3 := \braket{u_a(k_3) | \sig_3 | u_a(k_3)}$ of the eigenstate $\ket{u_a(k_3)}$. 
Due to the flux $\phi$, the Landau gap with size $2 \omega_L := 2\sqrt{2B_s} \approx 1.6$ opens near the Weyl point at $k_3 = 0$, and a spin-polarized chiral fermion emerges inside the gap.

The chiral dispersion emerging under a magnetic field produces a current parallel to the field, leading to a Floquet realization of the CME. 
To see this, we calculate the amount of charge $\Delta Q$ pumped during one period using $\Delta Q := \int_0^T dt J_3(t)$, where $J_3(t):=\int_{-\pi}^\pi {dk_3 \over 2\pi}\mr{Tr}\rbr{\rho_t \partial_{k_3} H(k_3, t)}$ is a current parallel to the magnetic field, $\rho_t$ is the density matrix at time $t$, and $H(k_3,t)$ is the time-dependent Hamiltonian. 
As shown in Ref.~\cite{Kitagawa10}, the pumped charge can be rewritten in terms of the quasienergy $\epsilon_{k_2,b}(k_3)$ of $\bar{U}(k_2,k_3)$ [$b$ specifies a Landau level] as 
\begin{align}
	\Delta Q = 
	\sum_{k_2, b}
	\int_{-\pi}^\pi {dk \over 2\pi} 
	f_{k_2, b}(k_3) {\partial \epsilon_{k_2, b}(k_3) \over \partial k_3}
	, \label{eq: 2}
\end{align}
where $f_{k_2, b}(k_3) := \bra{k_3, k_2, b}\rho_0 \ket{k_3, k_2, b}$ is a distribution function of the Floquet eigenstates $\ket{k_3, k_2, b}$ in the initial state. 
As an initial state, we here take an equilibrium state under a Hamiltonian $H_0 = \sum_{\bm{q}, \alpha}\epsilon_0(\bm{q})c_{\bm{q},\alpha}^\dagger c_{\bm{q},\alpha}$ on the lattice $L_C$ with dispersion relation $\epsilon_0(\bm{q}) := -t_3\cos(q_3/2)-t_1\cos q_1-t_2\cos q_2$ [$t_i>0$ is the hopping amplitude along the $x_i$ direction, see Fig.~\ref{fig: DynaWmag1} (d)]: 
\begin{align}
	\rho_0 &= \sum_{\bm{q}, \alpha = \uparrow, \downarrow} 
	f_{\mr{FD}}(\bm{q}) 
	\ket{\bm{q},\alpha}_0 \bra{\bm{q},\alpha}_0
	, 
	\label{eq: fcme, 8}
\end{align}
where $f_{\mr{FD}}(\bm{q})$ is the Fermi-Dirac distribution function with Fermi energy $E_F$, and $\ket{\bm{q},\alpha}_0$ is the Bloch state with momentum $\bm{q} \ (\in \rbr{-\pi,\pi}^2\times \rbr{-2\pi, 2\pi})$ and spin $\alpha$. 
Note that this thermal Fermi gas was used for detecting a static topological phase \cite{DauphinA13, RoyS16, RoyS18}. 
By tuning the lattice parameters to be $t_1,t_2 \ll t_3$, we can make the distribution $f_{k_2, b}(k_3)$ concentrate almost on the lower Floquet bands, as shown below.

In Fig.~\ref{fig: DynaWmag1} (c), we show the calculated pumped charge $\Delta Q$ for a half-filling and zero-temperature initial state with $t_1 = t_2 = t_3/3$. 
The obtained value close to $\phi/2$ [green dashed line in Fig.~\ref{fig: DynaWmag1} (c)] actually has a topological origin. 
In the limit $t_1, t_2\ll t_3$, only the lower Floquet bands are occupied, i.e., $f_{\mr{FD}}(\bk) = 1$, and hence 
\begin{align}
	f_{k_2, b}(k_3) = 
	\sum_{\bm{q}, \alpha} 
	|\braket{k_3, k_2, b|\bm{q}, \alpha}|^2
	= 1
	. 
\end{align}
Then, $\Delta Q$ in Eq.~\eqref{eq: 2} reduces to the sum of the one-dimensional winding number divided by the number of Landau levels $N_L = 2L_1$ [$L_1$ is the number of sites along the $x_1$ direction]. 
Since we have the $(\phi L_1)$-chiral bands due to the flux $\phi$, each of which has the winding number $+1$, we obtain $\Delta Q = \phi L_1 /N_L = \phi / 2$. 
We emphasize that this quantized chiral current cannot arise in usual Floquet-Weyl semimetals with topologically trivial $U(\bk)$ because left- and right-handed Weyl fermions appear with equal numbers in accordance with the Nielsen-Ninomiya theorem. 
The quasienergy band in our setup, in contrast, hosts a single-chirality Weyl fermion without a partner of opposite chirality within a single band, enabling us to realize the maximally imbalanced population with only one chiral component being occupied. 
Although the CME is a many-body phenomena originating from the chiral imbalance, a similar effect can be observed in the single-particle dynamics \cite{single_particle}. 
 
The chiral current in the Floquet CME is robust against perturbations of the model due to topological stability of the single Weyl point protected by the 3D winding number $W$. 
The blue points in Fig.~\ref{fig: DynaWmag1} (c) show the pumped charge under a modified protocol with imperfect spin-selective Thouless pumps, where $\mathcal{U}_j^\pm$ ($j=1,2,3$) is replaced by with $\mathcal{U}_j^\pm \mr{e}^{-i \delta \sigma_l}$ ($l=j+1$ mod 3). 
Although $\mathcal{U}_j^\pm$ is no longer spin-preserving and spin-selective due to the spin-mixing term $\mr{e}^{-i \delta \sigma_l}$, $\Delta Q$ is almost unaffected. 
Furthermore, we confirm that $\Delta Q$ persists even in finite temperature as shown by the orange points in Fig.~\ref{fig: DynaWmag1} (c).

\textit{Classification of gapless Floquet states.}--\ 
In static topological insulators, symmetries dramatically enrich the classification of insulators as well as that of their gapless surface states, which cannot be realized in bulk lattices under the symmetry constraint \cite{QiZhang11}. 
This fact naturally motivates us to topologically classify Floquet unitary operators under various symmetries. 
As shown below, the symmetry-protected Floquet unitaries offer a wide range of lattice-prohibited band structures under symmetries which include the single Weyl fermion with the topological Floquet unitary as an example. 

Let us take a Floquet operator $U(\bm{k})\in$ U($N$) given by some unitary matrix. We here consider three symmetries in the Altland-Zirnbauer classes \cite{Kitaev09, Schnyder08}: time-reversal symmetry $\Theta H(\bm{k},t)\Theta^{-1}=H(-\bm{k},T-t)$, particle-hole symmetry $CH(\bm{k},t)C^{-1}=-H(-\bm{k},t)$, and chiral symmetry $\Gamma H(\bm{k},t)\Gamma^{-1}=-H(\bm{k},T-t)$. In terms of the Floquet operators, these symmetries are expressed as $\Theta U(\bm{k})\Theta^{-1}=U^\dag(-\bm{k})$, $CU(\bm{k})C^{-1}=U(-\bm{k})$, and $\Gamma U(\bm{k})\Gamma^{-1}=U^\dag(\bm{k})$ \cite{Kitagawa10}. 
We allow any continuous deformation of Floquet operators which respect the symmetry of the system, and classify their stable equivalence classes according to the K-theory \cite{Kitaev09, TeoKane10}. Note that we do \textit{not} assume energy gaps of the quasienergy band. 
Then, the classification of the unitary matrices can be performed in a manner similar to the classification of ``unitary loops" for Floquet TIs/TSCs \cite{RoyHarper17, MorimotoPoVishwanath17}. 
Since the derivation is parallel to Refs.~\cite{RoyHarper17, MorimotoPoVishwanath17}, we here outline the general idea and give the full derivation in Supplemental Material \cite{derivation_classification}. 
We define a Hermitian matrix $H_U(\bm{k})$ by
\begin{equation}
H_U(\bm{k})=
\begin{pmatrix}
0 & U(\bm{k})\\
U^\dag(\bm{k}) & 0
\end{pmatrix},
\label{eq_HU}
\end{equation}
which satisfies $H_U(\bm{k})^2=\bm{1}_{2N}$ and thus has eigenvalues $\pm 1$. 
Note that in the case of the classification of Floquet TIs/TSCs, the unitary matrix is taken as $U(\bm{k},t):=\mathcal{T}\exp[-i\int_0^tdt'H(\bm{k},t')]$ instead of $U(\bm{k})$ \cite{RoyHarper17}. Regarding $H_U(\bm{k})$ as a ``Hamiltonian" and identifying its symmetry class, 
we can show that the classification of $U(\bm{k})$ is equivalent to that of $H_U(\bm{k})$ given by some K-groups of static TIs/TSCs.
Using the K-group isomorphism between different spatial dimensions \cite{Kitaev09, TeoKane10}, 
we find that the K-group of Floquet operators of a symmetry class in $d$ dimensions is given by that of static TIs/TSCs of the same symmetry class in $(d+1)$ dimensions. 
Since the latter is equivalent to the classification of $d$-dimensional gapless surface states through the bulk-boundary correspondence, we arrive at the conclusion that the classification of $d$-dimensional gapless Floquet states is equivalent to that of $d$-dimensional gapless surface states of TIs/TSCs. 
The final result is summarized in Table \ref{tab_AZ_maintext}. 
This result strongly suggests that the gapless surface states of TIs/TSCs, which cannot have any pure lattice realization without bulk, can be realized in bulk quasienergy spectra of periodically driven lattice systems. 
In fact, a single Weyl fermion presented in this paper corresponds to a surface state of a four-dimensional topological insulator \cite{Qi08} and to class A in $d=3$ in Table \ref{tab_AZ_maintext}.
It merits further study to explicitly construct examples of Floquet operators in other symmetry classes. 

\begin{center}
\begin{table}
\caption{Tenfold-way topological classification of Floquet operators for spatial dimension $d = 0,1,2,3$. The Floquet single Weyl fermion in Eq.~\eqref{eq: strobo0} corresponds to class A in $d=3$.}
\label{tab_AZ_maintext}
\begin{ruledtabular}
\begin{tabular}{lcccc}
class & $d=0$ & $d=1$ & $d=2$ & $d=3$ \\ \hline
A & $0$ & $\mathbb{Z}$ & $0$ & $\mathbb{Z}$ \\
AIII & $\mathbb{Z}$ & $0$ & $\mathbb{Z}$ & $0$ \\ \hline
AI & $0$ & $0$ & $0$ & $2\mathbb{Z}$ \\
BDI & $\mathbb{Z}$ & $0$ & $0$ & $0$ \\
D & $\mathbb{Z}_2$ & $\mathbb{Z}$ & $0$ & $0$ \\
DIII & $\mathbb{Z}_2$ & $\mathbb{Z}_2$ & $\mathbb{Z}$ & $0$ \\
AII & $0$ & $\mathbb{Z}_2$ & $\mathbb{Z}_2$ & $\mathbb{Z}$ \\
CII & $2\mathbb{Z}$ & $0$ & $\mathbb{Z}_2$ & $\mathbb{Z}_2$ \\
C & $0$ & $2\mathbb{Z}$ & $0$ & $\mathbb{Z}_2$ \\
CI & $0$ & $0$ & $2\mathbb{Z}$ & $0$ \\
\end{tabular}
\end{ruledtabular}
\end{table}
\end{center}

\textit{Summary.}--\ 
We have presented a periodically driven 3D lattice system that exhibits the single Weyl fermion in a quasienergy spectrum, thereby demonstrating a Floquet version of the CME. 
Our proposal utilizes the topology of the Floquet unitary. 
While the mathematical formula of the 3D winding number \eqref{eq: windnumber} was presented in a seminal work \cite{Kitagawa10}, its physical consequence and concrete realization had remained elusive. 
We have resolved this problem and provided a generalization to a topological classification of Floquet operators in the Altland-Zirnbauer symmetry classes. 
We expect that the unique topological structure arising from unitary operators will serve as a useful guideline for designing non-equilibrium systems free from the limitations of static phases of matter.

\begin{acknowledgments}
We thank I.\ Danshita, T.\ Tomita, Y.\ Ashida, Z.\ Gong, Y.\  Kikuchi, and T.\ Morimoto for fruitful discussions. 
We are especially grateful to K.\ Shiozaki for suggesting the generalization of our result by including symmetries. 
This work was supported by KAKENHI Grant No. JP18H01145 and a Grant-in-Aid for Scientific Research on Innovative Areas ``Topological Materials Science" (KAKENHI Grant No. JP15H05855) from the Japan Society for the Promotion of Science (JSPS). 
S. H. acknowledges support from JSPS (Grant No. JP16J03619) and through the Advanced Leading Graduate Course for Photon Science (ALPS). 
M.N. is supported by RIKEN Special Postdoctoral Researcher Program.
\end{acknowledgments}

\appendix

\section{\label{sec: 1D}Floquet realization of a single chiral fermion in a one-dimensional lattice}
A prototypical example of a gapless Floquet spectrum with nontrivial topology is a single chiral fermion, which is prohibited in a static one-dimensional lattice \cite{NielsenNinomiya1, NielsenNinomiya2}. 
This example corresponds to class A in $d=1$ in the classification table [see Table I in the main text]. 
The model is given by the Thouless pump \cite{Kitagawa10, Thouless83}, where an adiabatic cycle of a one-dimensional insulator leads to quantized charge pumping. 
Because of the adiabatic condition, 
an occupied state with momentum $k$ returns to the same state during one cycle; thus the Floquet operator is decomposed according to the momentum $k$ and the Floquet-band index $\alpha$: 
\begin{align}
	U(T) = \sum_{k, \alpha = 1,2} U_{F, \alpha}(k)
	. 
\end{align}
The two Floquet bands $\alpha = 1$ and $2$ correspond to the occupied and unoccupied bands of the initial Hamiltonian $H(t=0)$, respectively. 
When the system starts from the ground state of $H(t=0)$, the pumped charge $\Delta Q$ is expressed in terms of the first Chern number of the time-dependent Bloch Hamiltonian \cite{Thouless83}. 
Furthermore, $\Delta Q$ is rewritten in terms of the Floquet operator $U_{F, \alpha=1}(k)$ acting on the lower Floquet band as follows \cite{Kitagawa10}:
\begin{align}
	\Delta Q & = 
	\int_{-\pi}^{\pi} {dk \over 2\pi} 
	\mr{Tr}[U_{F, \alpha=1}^\dagger(k) i \partial_k U_{F, \alpha=1}(k)] 
	\nonumber \\
	 & = T \int_{-\pi}^{\pi} {dk \over 2\pi} 
	 {\partial \epsilon_{\alpha=1}(k) \over \partial k}
	 =: \nu_{\alpha=1}
	, \label{eq: fl, 2}
\end{align}
where $\nu_{\alpha=1}$ is the winding number of the quasienergy band over the Brillouin zone. 
For a unit winding $\nu_1=1$, $\epsilon(k)$ is topologically equivalent to $\epsilon(k) = k/T$, i.e., the band with a single chiral fermion. 
From the $(2\pi/T)$-periodicity of the quasienergy, this linear dispersion $\epsilon(k) = k/T$ is actually \textit{continuous} at $k=\pm\pi$. 
Thanks to the periodic structure in the quasienergy space, 
Floquet systems can realize topological band structures that are not achievable in static systems.

\section{\label{sec: smash}Nontriviality of $U(\bk)$ as a map from $\mbT^3$ to $\mr{SU}(2)$} 
It follows from the isomorphism $\mr{SU}(2) \cong S^3$ that the element $U$ in $\mr{SU}(2)$ can be parametrized in terms of $\bm{u} = \br{u_1, u_2, u_3, u_4} \in S^3$ as 
\peqn{
&U = u_4 \sig_0 + i \br{
u_1 \sig_1 + u_2 \sig_2 + u_3 \sig_3
}
, \nonumber \\
&\sum_{k = 1}^4 (u_k)^2 = 1
}
If we parametrize $U(\bk)$ as 
\ceqn{
U(\bk) = u_4(\bk) \sig_0 + i \br{
u_1(\bk) \sig_1 + u_2(\bk) \sig_2 + u_3(\bk) \sig_3
}
}
the winding number $W$ defined in Eq.~(6) in the main text can be expressed as 
\ceqn{
W = \int {d \bm{k} \over 2 \pi^2} 
\sum_{i,j,k,l = 1}^4 \epsilon_{ijkl} 
u_i(\bk)
{\partial u_j(\bk) \over \partial k_1}
{\partial u_k(\bk) \over \partial k_2}
{\partial u_l(\bk) \over \partial k_3}
\label{eq: wind, another}
}
where $\epsilon_{ijkl}$ is the totally antisymmetric tensor of rank $4$ \cite{MantonN04}. 
Expanding Eq.~(5) in the main text, 
we obtain 
\peqn{
u_1(\bk) =& 
- \sin\br{k_1} 
\cos^2\br{{k_2 \over 2}}
\cos^2\br{{k_3 \over 2}}
, \nnco
u_2(\bk) =& 
- \cos^2\br{{k_1 \over 2}}
\sin(k_2)
\cos\br{{k_3 \over 2}}
\nnco
& + {1 \over 2} \sin(k_1)
\cos^2\br{{k_2 \over 2}}
\sin(k_3)
, \nnco
u_3(\bk) =& 
- {1 \over 2}
\sin(k_1) 
\sin(k_2)
\cos\br{{k_3 \over 2}}
\nnco
&- \cos^2\br{{k_1 \over 2}}
\cos^2\br{{k_2 \over 2}}
\sin(k_3)
, \nnco
u_4(\bk) =& 
2 \cos^2\br{{k_1 \over 2}} 
\cos^2\br{{k_2 \over 2}}
\cos^2\br{{k_3 \over 2}} - 1
\label{eq: Uk, as, S3}
}
Substituting these into Eq.~\eqref{eq: wind, another}, we obtain $W = 1$. 

In general, a nontrivial map from $\mbT^3$ to $S^3$ can be constructed from the \textit{smash product} \cite{Nakahara03}. 
It is a mathematical tool for constructing a manifold from two manifolds. 
Let $X$ ($Y$) be a manifold and $x_0$ ($y_0$) be a point on $X$ ($Y$). 
The smash product $X \wedge Y$ is defined as the product space $X \times Y$ with the space $(\tbr{x_0} \times Y) \cup (X \times \tbr{y_0} )$ identified with a point on $X \times Y$. 
For example, the smash product $S^1 \wedge S^1$, with $S^1$ being a circle, is isomorphic to the two-dimensional sphere $S^2$. 
We parametrize these two circles as 
\ceqn{
S^1 = \tbr{k | -\pi \le k \le \pi}, \ 
S^1 = \tbr{k\pri | -\pi \le k\pri \le \pi}
}
and take $x_0$ and $y_0$ as $k = \pm\pi$ and $k\pri = \pm\pi$, respectively. 
Then, the product space $S^1 \times S^1$ and its subspace, 
\ceqn{
&(\tbr{x_0} \times Y) \cup (X \times \tbr{y_0} )
 = 
 \nnco
&\tbr{(\pm\pi, k\pri)  | -\pi \le k\pri \le \pi} \cup \tbr{(k, \pm\pi) | -\pi \le k \le \pi}
\label{eq: smash11boundary}
}
are identified with the Brillouin zone $\mbT^2$ of a square lattice and its boundary, respectively. 
As we can obtain $S^2$ from $\mbT^2$ by wrapping up the square $\rbr{-\pi, \pi}^2$, 
the smash product $S^1 \wedge S^1$, which is defined as the square $\rbr{-\pi, \pi}^2$ with its boundary identified with a point, is isomorphic to $S^2$. 
The isomorphic mapping $f_{1,1}$ is given by 
\ceqn{
& f_{1,1}(k, k\pri) 
\nnco
:= &
\rbr{- 2 \cos^2\br{{k \over 2}} \cos^2\br{{k\pri \over 2}} + 1 }\bm{n}_1
\nnco
&- \cos^2\br{{k \over 2}} \sin\br{k\pri} \bm{n}_2
 - \sin\br{k} \cos\br{{k\pri \over 2}} \bm{n}_3
\label{eq: isomorphism11}
}
where $\bm{n}_1 = (1,0,0)^T, \bm{n}_2 = (0,1,0)^T$, and $\bm{n}_3 = (0,0,1)^T$ (the superscript $T$ denotes the transpose) are the unit vectors in three orthogonal directions. 

As shown below, 
the nontrivial map $U(\bk)$ is constructed from the following isomorphism \cite{Nakahara03}: 
\peqn{
S^1 \wedge S^1 \wedge S^1 \cong 
S^2 \wedge S^1 \cong 
S^3 
\label{eq: double, smash}
}
Let $\bm{\xi} = (\xi_1, \xi_2, \xi_3)^T$ be a unit vector on $S^2$, i.e., $\sum_{i=1}^3 (\xi_i)^2 = 1$, and $k_1 \in [-\pi, \pi] \cong S^1$. 
The isomorphic mapping $f\colon S^2 \times S^1 \to S^3$ of the second isomorphism in Eq.~\eqref{eq: double, smash} is given by 
\ceqn{
&f_{2,1}(\bm{\xi}, k_1) 
\nnco
=& {\xi_1 - 1 \over 2} \sin\br{k_1} \bm{a}_1 
 + \cos\br{{k_1 \over 2}} 
 \br{\xi_3 \bm{a}_2 + \xi_2 \bm{a}_3}
\nnco
& - \rbr{ \sin^2\br{{k_1 \over 2}} + \cos^2\br{{k_1 \over 2}} \xi_1} \bm{a}_4
\label{eq: isomorphism21}
}
where $\bm{a}_i (i=1,2,3,4)$ are the unit vectors in four dimensions defined by 
\peqn{
\bm{a}_1 &= (1,0,0,0)^T, \quad \bm{a}_2 = (0,1,0,0)^T
, \nnco
\bm{a}_3 &= (0,0,1,0)^T, \quad \bm{a}_4 = (0,0,0,1)^T
}
Then, the composition of $f_{1,1}$ and $f_{2,1}$ defined by 
\neqn{
\widetilde{\bm{u}}(\bk) = 
f_{2,1}\rbr{f_{1,1}(k_2, k_3), k_1}
}
gives an isomorphic mapping between $S^1 \wedge S^1 \wedge S^1$ and $S^3$. 
Since the isomorphic mapping $f_{2,1}\rbr{f_{1,1}(k_2, k_3), k_1}$ to $S^3$ naturally has a unit winding number and the domain of $S^1 \wedge S^1 \wedge S^1$ and that of $\mbT^3$ are both cuboid $\rbr{-\pi, \pi}^3$, 
$\widetilde{\bm{u}}(\bk)$ has a unit winding number: 
\peqn{
W &= \int {d \bm{k} \over 2 \pi^2} 
\sum_{i,j,k,l = 1}^4 \epsilon_{ijkl} 
\widetilde{u}_i(\bk)
{\partial \widetilde{u}_j(\bk) \over \partial k_1}
{\partial \widetilde{u}_k(\bk) \over \partial k_2}
{\partial \widetilde{u}_l(\bk) \over \partial k_3}
\nnco
 &= 1
}
We note that $\widetilde{\bm{u}}(\bk)$ stays constant on the boundary of the Brillouin zone $\mbT^3 = [-\pi,\pi]^3$: 
\peqn{
\widetilde{\bm{u}}(\bk) = \br{0,0,0,-1}
\label{eq: bc, appA}
}
We define a continuous deformation $\widetilde{\bm{u}}_s(\bk) := R_{23}(s k_1 / 2) \widetilde{\bm{u}}(\bk)$ with a deformation parameter $s (\in [0,1])$, where $R_{23}(\theta)$ is a rotation matrix defined by 
\peqn{
R_{23}(\theta) := \br{\begin{array}{cccc}
	1&0&0&0 \\
	0&\cos\theta & -\sin\theta&0 \\
	0&\sin\theta &\cos\theta&0 \\
	0&0&0&1 
\end{array}}
}
It follows from Eq.~\eqref{eq: bc, appA} that 
$\widetilde{\bm{u}}_s(\bk) = \br{0,0,0,-1}$ on the boundary of $\mbT^3$
and hence $\widetilde{\bm{u}}_s(\bk)$ is continuous on $\mbT^3$. 
Also, combining Eqs.~\eqref{eq: Uk, as, S3}, \eqref{eq: isomorphism11}, and \eqref{eq: isomorphism21}, we obtain 
\ceqn{
\widetilde{\bm{u}}_{s=0}(\bk) &= \widetilde{\bm{u}}(\bk)
, \nnco
\widetilde{\bm{u}}_{s=1}(\bk) &= \bm{u}(\bk)
}
where $\bm{u}(\bk)$ is defined in Eq.~\eqref{eq: Uk, as, S3}. 
Therefore, $\bm{u}(\bk)$ is also an isomorphic mapping between $S^1 \wedge S^1 \wedge S^1$ and $S^3$ and has a unit winding number $W=1$.

\section{\label{sec: experiment}Experimental implementation with ultracold atomic gases}
In this appendix, 
we discuss possible experimental implementations of the Floquet CME either by using a spin-dependent optical lattice with ${}^{87}$Rb or by using laser-assisted tunneling with ${}^{173}$Yb. 

\subsection{Setup with ${}^{87}$Rb}
\begin{figure}[t]
  \begin{center}
	\includegraphics[clip, width = \columnwidth]{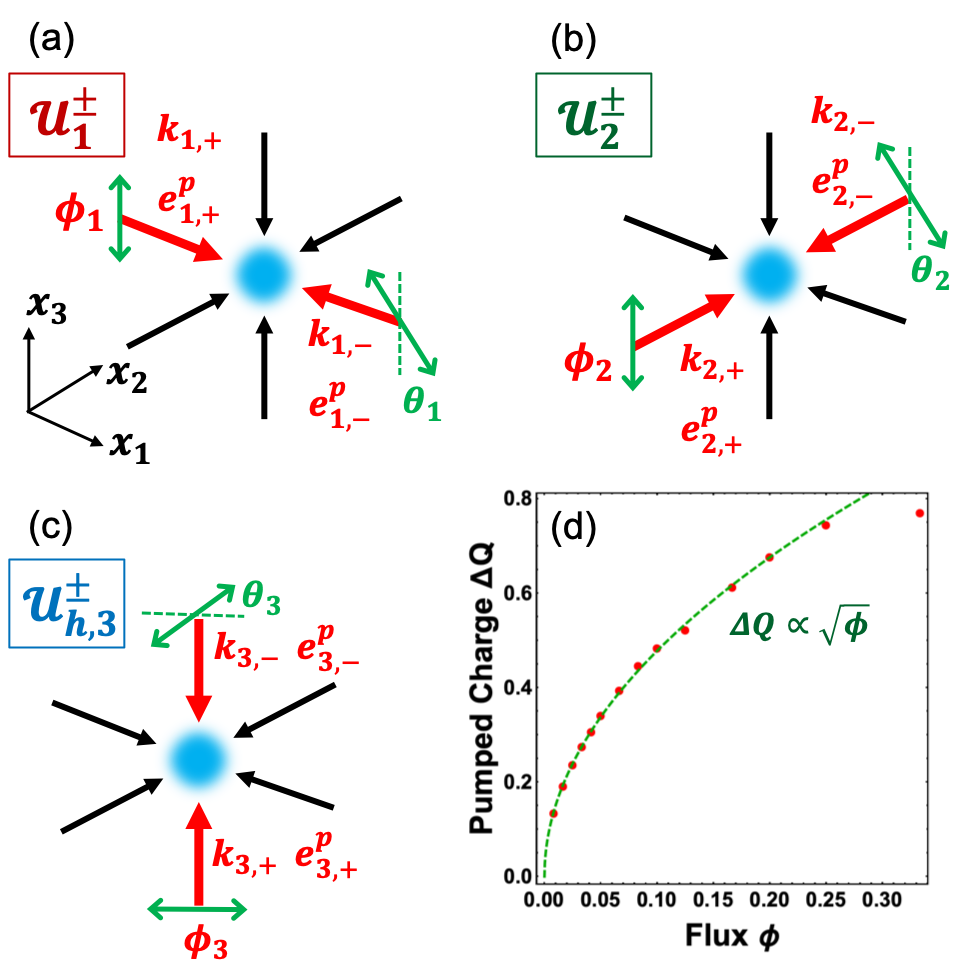}
	\caption{
(a), (b), (c) Laser configurations of spin-selective pumps 
(a) $\mathcal{U}_1^\pm$, (b) $\mathcal{U}_2^\pm$, and (c) $\mathcal{U}_{h,3}^\pm$, where $\bk_{i,\alpha}$ and $\bm{e}^p_{i, \alpha} \ (i = 1,2,3$ and $\alpha = +, -)$ are the wave vectors and the polarization vectors of the six linearly polarized laser beams forming an optical lattice. 
The red and black arrows represent the wave vectors of the counterpropagating laser beams and green arrows are their polarization directions, and $\theta_i$ denotes the angle between $\bm{e}^p_{i,+}$ and $\bm{e}^p_{i,-}$, and $\phi_i$ denotes the locked phase of the incoming wave along the $x_i$ axis. 
(d) Pumped charge $\Delta Q$ for a bosonic atomic gas under the magnetic flux $\phi$, where the initial state is taken as a Bose-Einstein condensate at the momentum $\bk=0$. 
	}
	\label{fig: experiment}
  \end{center}
\end{figure}

As discussed in the main text, the main ingredient of our driving protocol 
is spin-dependent transport, which has already been realized with ${}^{87}$Rb atoms in spin-dependent optical lattices \cite{Mandel03a, Mandel03b}. 
The two spin states $\left| \uparrow \right>$ and $\left| \downarrow \right>$ are chosen as $\left| \uparrow \right> = \left| F=2, m_F = -2 \right>$ and $\left| \downarrow \right> = \left| F = 1, m_F = -1 \right>$. 
Consider a 3D optical lattice produced by three orthogonal laser beams and their retroreflected ones. 
To suppress the natural hopping $J$, we apply an optical  field gradient with a slope $\Delta$, which is sufficiently small compared with the lattice depth $V_0$ but larger than $J$ \cite{JakschD03, AidelsburgerM14}. 
The three pairs of counterpropagating plane waves are all linearly polarized, 
and we write the angle between the two polarization vectors $\bm{e}^p_{1, +}$ and $\bm{e}^p_{1,-}$ of the beams along the $x_1$ axis as $\theta_1$, that along the $x_2$ axis as $\theta_2$, and that along the $x_3$ axis as $\theta_3$. 
Those angles are dynamically controlled by electro-optical modulators by rotating the polarization vector of the retroreflected laser beams. 
Then, the wave vectors $\bk_{i,\alpha} \ (i = 1,2,3$ and $\alpha = +, -)$ and the polarization vectors $\bm{e}^p_{i,\alpha} \ (i = 1,2,3$ and $\alpha = +, -)$ of the six laser beams are given by 
\begin{align}
	& \bm{k}_{1, \pm} = \pm k \bm{n}_1, 
	\ \bm{e}^p_{1, \pm}, = R_1\br{{\pm \theta_1 \over 2}} \bm{n}_2, 
	\nnco
	& \bm{k}_{2, \pm} = \pm k \bm{n}_2, 
	\ \bm{e}^p_{2, \pm} = R_2\br{{\pm \theta_2 \over 2}} \bm{n}_3, 
	\nnco
	& \bm{k}_{3, \pm} = \pm k \bm{n}_3, 
	\ \bm{e}^p_{3, \pm} = R_3\br{{\pm \theta_3 \over 2}} \bm{n}_1
	, 
\end{align}
where $\bm{n}_i \ (i = 1,2,3)$ is the unit vector in the $x_i$ axis and $R_i(\theta)$ is the rotation matrix around the $x_i$ axis through angle $\theta$ [see Figs.~\ref{fig: experiment} (a), (b), and (c)]. 

Then, the created optical potential $U(\bm{x})$ at position $\bm{x}$ is given by 
\begin{align}
	U(\bm{x}) &= u_s |\rbr{\bm{e}^p(\bm{x})}^\ast \cdot \bm{e}^p(\bm{x})| \sigma_0 
	+ u_v \bm{B}_{\mr{eff}}\cdot \bm{\sigma}
	, \nonumber \\
	\bm{B}_{\mr{eff}} &= i \rbr{\bm{e}^p(\bm{x})}^\ast \times \bm{e}^p(\bm{x})
	, \nonumber \\
	\bm{e}^p(\bm{x}) &= \sum_{i=1,2,3}\sum_{\alpha = \pm} 
	\bm{e}^p_{i, \alpha} \exp(i \bm{k}_{i, \alpha} \cdot \bm{x} + \phi_i)
	, 
\end{align}
where $u_s$, $u_v$, $\sigma_0$, and $\bm{\sigma}$ are the scalar potential, the vector potential, the $2 \times 2$ identity matrix, and the Pauli matrix vector with respect to the spin states $\left| \uparrow \right>$ and $\left| \downarrow \right>$, respectively \cite{Goldman14b}. 
Here, $\phi_i$ is the phase of the incoming wave along the $x_i$ axis that is phase-locked and controlled by an additional electro-optical modulator. 
We first consider the spin transport along the $x_1$-axis through the changes of $\theta_1$ and $\phi_1$. 
Suppose $u_s > u_v > 0$ and we initially set $\theta_1 = \theta_2 = \theta_3 = 0$ and $\phi_1 = \phi_2 = \phi_3 = \pi$. 
In this case, atoms are trapped at $\bm{x} = a_{\mr{lat}}(m_1, m_2, m_3) \in a \mathbb{Z}^3$, where $a_{\mr{lat}} = \pi/k$ is the lattice constant.   
As we change $(\theta_1, \phi_1)$ from $(0, \pi)$ to $(\pi, 2\pi)$, atoms with spin states $\sig_1 = 1$ are displaced by one lattice site in the positive $x_1$ direction, while those with spin states $\sig_1 = -1$ are not, realizing spin-selective transport for spin states $\sig_1 = 1$. 
For the parameter change of $(\theta_1, \phi_1)$ from $(0, \pi)$ to $(\pi, 0)$, atoms with spin states $\sig_1 = -1$ are displaced by one lattice site in the negative $x_1$ direction, while those with spin states $\sig_1 = 1$ are not, realizing a spin-selective transport for spin states $\sig_1 = -1$. 
The spin-selective transport along the $x_2$- and $x_3$-directions can be achieved by changing $(\theta_2, \phi_2)$ and $(\theta_3, \phi_3)$, respectively.

Although ${}^{87}$Rb is a boson and hence the pumped charge $\Delta Q$ is different from the fermionic case shown in Fig.~2 (c) in the main text, 
we still have a nonzero $\Delta Q$ originating from the chiral dispersion in the quasienergy in Fig.~2 (b) in the main text. 
In Fig.~\ref{fig: experiment} (d), we calculate $\Delta Q$ [red points], where the initial state is taken as Bose-Einstein condensates at $\bk=0$. 
At zero temperature, the initial-state density matrix $\rho_0$ of non-interacting bosons is given by the mixture of Bose-Einstein condensates with spin up and down with momentum $\bk=0$: 
\begin{align}
	\rho_0 = {1 \over 2}\sum_{\alpha = \uparrow, \downarrow}\ket{\bk=0,\alpha}\bra{\bk=0,\alpha}
	. \label{eq: initial_state, BEC}
\end{align}
As shown by the red points in Fig.~\ref{fig: experiment} (d), $\Delta Q$ exhibits the power-law behavior $\Delta Q \propto \sqrt{\phi}$ for a small $\phi$ [green dashed curve], which is derived as follows. 
From Eq.~(7) in the main text and Eq.~\eqref{eq: initial_state, BEC}, we have 
\begin{align}
	\Delta Q = 
	\sum_{b}
	|\langle k_2 = k_3 = 0, b | \bm{k} = 0 \rangle |^2 {\partial \epsilon_{k_2, b}(k_3=0) \over \partial k_3}
	, 
\end{align}
Since we have the $(\phi L_1)$ chiral bands and the overlap integral satisfies $|\langle k_2 = k_3 = 0, b | \bm{k} = 0 \rangle |^2 \propto (\sqrt{\phi}L_1)^{-1}$, we obtain $\Delta Q \propto (\phi L_1) \times (\sqrt{\phi}L_1)^{-1} \propto \sqrt{\phi}$.

To observe of the chiral transport in Fig.~\ref{fig: experiment} (d), the time scale of the pump $\tau_{\mr{pump}}$ should be made much smaller than the decoherence time $\tau_{\mr{de}}$. 
The time scale of the pump $\tau_{\mr{pump}}$ is determined from the adiabatic condition of sliding of the lattice, i.e., avoiding the excitation to higher Bloch bands. 
This condition is satisfied for $\tau_{\mr{pump}} \ge 40 \ \mr{\mu}$sec for the lattice depth $V_0 = 30 E_r$, with $E_r$ being the recoil energy \cite{Mandel03a, Mandel03b}. 
The dominant mechanism for decoherence may be the on-site interaction between particles \cite{Mandel03a}, by which $\tau_{\mr{de}} = $ 200 $\mu$sec for the lattice depth $V_0 = 25 E_r$ with $E_r$ being the recoil energy of ${}^{87}$Rb with wavelength $\lambda = 785$ nm. 
When the single spin-selective Thouless pump operates within $40 \ \mr{\mu}$sec, atoms experience the interaction energy during the time $\tau_{d,s} = 40 \ \mr{\mu}\mr{sec} \times (w/\lambda)$ with $w$ being the size of a wave packet localized at a site. 
Then, the total time $\tau_{d,\mr{tot}}$ during which the atoms experience the interaction is estimated to be $\tau_{d,\mr{tot}} = 8 \tau_{d,s}$. 
Since $(w/\lambda) \approx \sqrt{V_0/E_r}$ for a deep optical lattice and the interaction energy $\propto \tau_{\mr{de}}^{-1}$ is proportional to $(V_0/E_r)^{\frac{4}{3}}$ \cite{Bloch08}, we have $\tau_{\mr{de}}/\tau_{d, \mr{tot}} = 8$ for the lattice depth $V_0 / E_r = 20$. 
In this case, the optical field gradient with the on-site energy difference $\Delta$ with tens of kHz is sufficient to suppress the natural hopping $J = 0.02 E_r \sim $ kHz.  
Thus, 8 cycles of the pumps can operate, which is sufficient to observe the displacement of the center of mass shown in Fig.~\ref{fig: experiment} (d). 
Another limitation is the excitation to higher Bloch bands due to the quadrupole field pulse. 
The weight $w_{\mr{exc}}$ of the excited states is $w_{\mr{exc}} \propto \phi (w/\lambda)^2$ \cite{Sorensen05} and hence it is negligible for a weak magnetic flux and a deep optical lattice.   

\subsection{Setup with ${}^{173}$Yb}
Motivated by the recent experiments on realizing synthetic gauge fields \cite{Mancini15, Song16} and the proposals for implementing a helical hopping \cite{BudichJ15, Budich17}, we consider yet another implementation scheme using fermionic alkaline-earth-like atoms ${}^{173}$Yb and a laser-assisted hopping using an excited level ${}^3$P${}_1$. 
For the implementation, we rewrite $U(\bk)$ as combination of helical pumps as follows: 
\begin{align}
	U(\bk) =& \wt{U}_1(k_1) \wt{U}_3\br{ {k_3 \over 2}} \wt{U}_2(k_2) \wt{U}_3\br{ {k_3 \over 2}} 
	\nonumber \\
	&\times \wt{U}_1(k_1) \wt{U}_3\br{ {k_3 \over 2}} \wt{U}_2(k_2) \wt{U}_3\br{ {k_3 \over 2}} 
	, 
\end{align}
where $\wt{U}_j(k) := \mr{e}^{- i\sig_j k/2}$ gives a unitary operation of the helical pump. 
The two spin states are taken as $\left| m_F = -5/2 \right>$ and $\left| m_F = -1/2 \right>$ in the ground-state manifold ${}^1$S${}_0$ with $F = I = 5/2$. 
The six-fold degeneracy between the spin states is lifted  by a Zeeman splitting $\Delta_{B, 3}$ induced by a uniform magnetic field in the $x_3$ direction. 
The natural hopping is suppressed by a magnetic field gradient $\delta B_3 = \Delta_1 x_1 + \Delta_2 x_2 + \Delta_3 x_3$, where $\Delta_1, \Delta_2$, and $\Delta_3$ are the on-site energy difference along the $x_1$, $x_2$, and $x_3$ directions, respectively. 
For a lattice depth of $V_0  = 20 E_r$, where the natural hopping $J$ is given by $J = 0.02 E_r$, the Zeeman splitting $\Delta_{B,z}$ of the order of tens of kHz is sufficient for the suppression. 
A spin-flip hopping along the $x_3$ direction can be induced by Raman laser beams which are resonant to the energy difference $\Delta_3$ \cite{BudichJ15}. 
The helical hoppings $\wt{U}_{1,2}(k_{1,2})$ in the other two directions can be implemented by a combination of the $\pi/2$ pulses and the helical pump because we have 
\begin{align}
	\wt{U}_1(k_1) = 
	\Exp{-i\pi \sig_2/ 4} \wt{U}_3(k_1) \Exp{i\pi \sig_2/ 4}
	, \nonumber \\
	\wt{U}_2(k_2) = 
	\Exp{i\pi \sig_1/ 4} \wt{U}_3(k_2) \Exp{-i\pi \sig_1/ 4}
	. 
\end{align}
One can selectively induce these three directional hoppings by making $\Delta_1, \Delta_2$, and $\Delta_3$  different from each other and three pairs of Raman laser beams. 
This also can be done by a pair of Raman laser beams and dynamically changing $\Delta_1, \Delta_2$, and $\Delta_3$. 

To observe the Floquet CME, the time scale of the pump $\tau_{\mr{pump}}$ should be made much smaller than the lifetime $\tau_{\mr{life}}$ of the Raman induced hopping. 
The time scale of a Raman-induced hopping $\Omega$ on a lattice system is of the order of kHz \cite{Mancini15} when the one-photon detuning $\delta$ of the Raman process is taken to be of the order of 1 GHz \cite{BudichJ15, Song16}. 
In this case, the single pump operates within one msec, leading to tens of msec for one cycle of the pump. 
The lifetime $\tau_{\mr{life}} \sim \delta / (\gamma \Omega)$ due to heating with the Raman process is of the order of 1 sec for the lifetime of $\gamma = 850$ nsec \cite{Enomoto08}, which is much longer than that of an alkali-metal system \cite{Song16}. 
Therefore, at least a few tens of pumps can operate within the lifetime, which is sufficient for the observation of the Floquet CME.

\section{\label{sec: explicit}Explicit form of $\bar{U}(k_2, k_3)$}
Let $L_1$ be the number of sites along the $x_1$ direction. 
The partial Fourier transforms of $\mathcal{U}_{h,3}^- \mathcal{U}_q^\dagger \mathcal{U}_2^\pm \mathcal{U}_q \mathcal{U}_{h,3}^+$ and $\mathcal{U}_1^{\pm}$ in the $x_2$- and $x_3$-axes are given by 
\begin{align}
	&\mathcal{U}_{h,3}^- \mathcal{U}_q^\dagger \mathcal{U}_2^\pm \mathcal{U}_q \mathcal{U}_{h,3}^+ 
	= \sum_{k_2, k_3} \sum_{x_1=1}^{L_1} 
	\bar{c}_{x_1}^\dagger \widetilde{U}_{k_2,k_3}^\pm(x_1) \bar{c}_{x_1}
	, \nonumber \\
	&\mathcal{U}_1^\pm = \sum_{k_2, k_3} \sum_{x_1=1}^{L_1}
	\br{
	\bar{c}_{x_1 \pm 1}^\dagger P_1^\pm \bar{c}_{x_1}
	 + \bar{c}_{x_1}^\dagger P_1^\mp \bar{c}_{x_1}
	}
	, 
\end{align}
where $\bar{c}_{x_1} := c_{x_1, k_2, k_3}$ is the annihilation operator of the fermion at site $x_1$ with fixed momenta $k_2$ and $k_3$ and $\widetilde{U}_{k_2,k_3}^\pm(x_1) := U_{h,3}^-(k_3)U_2^\pm(k_2 - 2 \pi \phi x_1)U_{h,3}^+(k_3)$. 
Since $k_2$ and $k_3$ remain good quantum numbers, the Floquet operator $\mathcal{U}_F$ acting on the lower Floquet bands is decomposed according to them: 
\begin{align}
	& \mathcal{U}_F =
	\mathcal{U}_1^- 
	\br{
	\mathcal{U}_{h,3}^- \mathcal{U}_q^\dagger \mathcal{U}_2^- \mathcal{U}_q \mathcal{U}_{h,3}^+
	}
	\mathcal{U}_1^+ 
	\br{
	\mathcal{U}_{h,3}^- \mathcal{U}_q^\dagger \mathcal{U}_2^+ \mathcal{U}_q \mathcal{U}_{h,3}^+
	}
	\nonumber \\ 
	&\quad 
	=: \sum_{k_2, k_3}U\pri(k_2, k_3)
	.  \label{eq: fcme, 25}
\end{align}
However, this naive introduction of the flux breaks the  the periodic boundary condition along the $k_3$ direction, i.e, $U\pri(k_2, \pi) \ne U\pri(k_2,-\pi)$, due to the coupling between the lower and higher Floquet bands. 
To completely decouple them, we introduce the additional time evolution $\mathcal{U}_s$ with duration $\tau_s$ generated by a static Hamiltonian 
\begin{align}
	H_s &= J_s \sum_{\bm{x},\alpha} (i c_{\bm{x} + {\bm{e}_3 \over 2},\alpha}c_{\bm{x},\alpha} + \mr{h.c.})
	\nonumber\\
	& = 2J_s\sum_{\bm{k},\alpha} \sin\br{{k_3 \over 2}}
	c_{\bm{k},\alpha}^\dagger c_{\bm{k},\alpha}
	, \label{eq: fcme, 20}
\end{align}
which can be implemented by laser-assisted hopping \cite{Goldman14b}. 
By tuning $2J_s\tau_s = \pi\phi/2$, we obtain the time-evolution operator $\bar{U}(k_2, k_3)$ acting only on the lower Floquet bands
\begin{align}
	\bar{U}(k_2, k_3) := U_{s,k_3} U\pri(k_2, k_3)
	, 
\end{align}
where $U_{s, k_3} := \exp\rbr{-i (\pi\phi/2) \sin(k_3/2)}$ is the Floquet operator acting on the lower Floquet bands. 
A straightforward calculation shows that $\bar{U}(k_2, k_3)$ satisfies the periodic boundary condition: $\bar{U}(k_2, \pi) = \bar{U}(k_2, -\pi)$. 
The overall Floquet operator $U(k_2, k_3)$ is given by 
\begin{align}
	\bar{U}(k_2, k_3) &= 
\sum_{x_1=1}^{L_1} 
\br{
\bar{c}_{x_1+1}^\dagger \bar{u}_1 \bar{c}_{x_1} + 
\bar{c}_{x_1}^\dagger \bar{u}_0 \bar{c}_{x_1} + 
\bar{c}_{x_1 - 1}^\dagger \bar{u}_{-1} \bar{c}_{x_1}
}
	, \nonumber \\
	\bar{u}_1 &= 
	U_{s, k_3} P_1^+ \widetilde{U}_{k_2,k_3}^-(x_1+1) P_1^+ \widetilde{U}_{k_2,k_3}^+(x_1)
	, \nonumber \\
	\bar{u}_0 &= 
	U_{s, k_3}P_1^- \widetilde{U}_{k_2,k_3}^-(x_1+1) P_1^+ \widetilde{U}_{k_2,k_3}^+(x_1) 
		\nonumber \\
		&\quad 
		+ U_{s, k_3}P_1^+ \widetilde{U}_{k_2,k_3}^-(x_1) P_1^- \widetilde{U}_{k_2,k_3}^+(x_1)
	, \nonumber \\
	\bar{u}_{-1} &= 
	U_{s, k_3} P_1^- \widetilde{U}_{k_2,k_3}^-(x_1) P_1^- \widetilde{U}_{k_2,k_3}^+(x_1)
	. 
\end{align}
The unitary operator $\bar{U}(k_2, k_3)$ gives a Floquet analog of the Aubry-Andrei-Harper model \cite{Aubry80, Harper55} with two parameters $k_2$ and $k_3$.

\begin{figure}[t]
  \begin{center}
	\includegraphics[clip, width = \columnwidth]{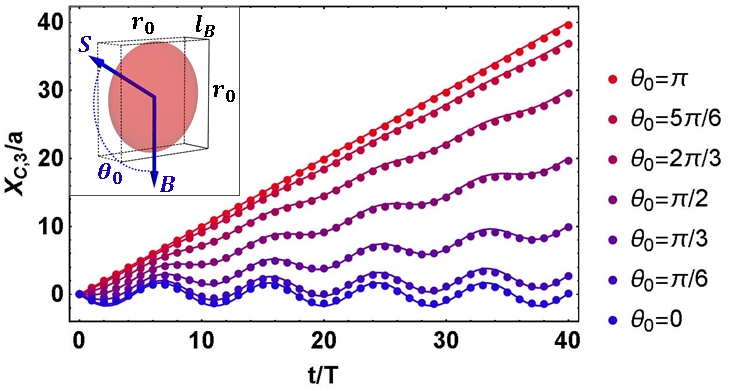}
	\caption{
Time evolution of the center of mass $X_{C, 3}$ along the direction of the applied magnetic field $\bm{B} = (0,0, -2 \pi \phi)$ with $\phi = 1/25$ for relative angles $\theta_0 = 0$ [blue], $\pi/6, \pi/3, \pi/2, 2\pi/3, 5\pi/6, \pi$ [red]. 
The solid curves are the theoretical ones in Eq.~\eqref{eq: zitter}. 
The inset shows a schematic illustration of the initial pancake-shaped wave packet with widths $l_B, r_0$, and $r_0$ in the $x_1$-, $x_2$-, and $x_3$-directions, respectively, and $\theta_0$ is the angle between $\bm{B}$ and the spin polarization $\bm{S}$. 
}
	\label{fig: single particle}
  \end{center}
\end{figure}
\section{\label{sec: single}Single-particle dynamics in a chiral Floquet band}
In this section, we consider the single-particle dynamics of a wave packet under the driving protocol $\bar{\mathcal{U}}_F$ in the main text, which is feasible for studying the Floquet CME in photonic Floquet systems \cite{GaoF16, MukherjeeS17, MaczewskyL17, SunX18}, quantum walks \cite{KitagawaT10, KitagawaT12}, and ultracold atomic gases \cite{JotzuG14, Stuhl15, RoyS18}. 

As we will see below, a spin-polarized wave packet moves in the direction opposite to the applied magnetic field under the driving due to the chiral dispersion. 
Furthermore, we find an oscillating motion of the wave packet originating from the \textit{Zitterbewegung} \cite{SchrodingerE30}. 
Let $l_B = 1/\sqrt{2 \pi \phi}$ be the magnetic length. 
For clear demonstration of the Floquet CME, we initially prepare a spin-polarized Gaussian wave packet $\psi(\bm{x}, t=0)$ with spin polarization $\bm{S}$ and  width $l_B (\gg a_{\mr{lat}})$ in the $x_1$ direction and $r_0 (\gg l_B)$ in the other directions [see the inset in Fig.~\ref{fig: single particle}], where 
\begin{align}
	\psi(\bm{x}, t=0) = 
\mathcal{N}^{-1/2}
\exp\rbr{-{ (x_1)^2 \over 2 (l_B)^2} - { (x_2)^2 + (x_3)^2 \over 2 (r_0)^2}}\bm{v}
	, 
\end{align}
with $\sqrt{\mathcal{N}} = (\pi)^{3/2} (l_B)^{1/2}r_0$. 
Here, $\bm{v}$ is a two-component vector representing the spin degrees of freedom: $\bm{S}=\left<\bm{v}, \bm{\sigma} \bm{v} \right>$. 
Then, the Fourier-transformed state $\psi(t=0, k_2, k_3)$ is localized at $k_2 = k_3 = 0$ and can be expanded as 
\begin{align}
	&\psi(t=0, k_2, k_3) 
	\nonumber \\
	&\propto
	\sin\br{{\theta_0 \over 2}} \Psi_0(k_3) + 
	i \cos\br{{\theta_0 \over 2}} 
	{\Psi_1(k_3) - \Psi_{-1}(k_3) \over \sqrt{2 \omega}} 
	, 
\end{align}
where $\Psi_{-1}(k_3), \Psi_0(k_3)$, and $\Psi_1(k_3)$ are the lowest particle band, the chiral band, and the highest hole band of the Landau levels, respectively, 
and $\theta_0$ is the angle between $\bm{B}$ and $\bm{S}$. 
The wave function at time $t$ is given by 
\begin{align}
	\bar{\psi}(t, k_3) \propto &
 \sin\br{{\theta_0 \over 2}} \mr{e}^{-i \phi_0 - i \epsilon_0 t} \Psi_0(k_3)
\nnco
+ & {i v  \cos\br{{\theta_0 \over 2}} \over \epsilon_1 l_B}
 \br{
{ \mr{e}^{- i \epsilon_1 t} \Psi_1(k_3) \over \sqrt{ \eps_1 + v k_3} }
 - { \mr{e}^{i \epsilon_1 t} \Psi_{-1}(k_3) \over \sqrt{ \eps_1 - v k_3} }
}, 
\end{align}
where $\epsilon_i$ is the quasienergy of the band $\Psi_i(k_3)$. 
For the Weyl equation in the continuous spacetime under a magnetic field $i \partial_t \psi = (\bk + \bm{A})\cdot \bm{\sigma} \psi$ with $\bm{A} = (0, - 2 \pi \phi x_1, 0)$, the time evolution of the center of mass $X_{C, 3}(t) = \int d\bm{x} \ x_3 |\psi(\bm{x}, t)|^2$ along the $x_3$ direction is given by 
\peqn{
& X_{C, 3}(t)
 = i \int dx_1 
\bar{\psi}^\ast(t, k_3 = 0) {\partial \bar{\psi}(t, k_3 = 0) \over \partial k_3} 
\nnco
& = \sin^2\br{ \theta_0 \over 2} v t
\nnco
&\quad - {i \over \sqrt{2} \omega} \cos^2\br{ \theta_0 \over 2} \rbr{
\mr{e}^{2 i \omega t} 
\int dx_1 \ 
\Psi_1^\ast 
{\partial \Psi_{-1}\over \partial k_3}  
+ \mr{c.c.}
}
\nnco
& = \sin^2\br{\theta_0 \over 2} v t - {v \over 2 \omega}
 \cos^2\br{\theta_0 \over 2} \sin(2 \omega t)
 \label{eq: zitter}
}
In the units $a_{\mr{lat}} = T = 1$, we have $v = a_{\mr{lat}}/T = 1$.

In Fig.~\ref{fig: single particle}, we numerically calculate $X_{C, 3}(t)$ at the discrete times $t = 0, T, 2T, \cdots$ based on the discrete-time Schr\"odinger equation $\psi(t + T) = \bar{\mathcal{U}}_F \psi(t)$ for various initial angle $\theta_0$. 
The numerical results [shown by points] are in excellent agreement with theoretical curve \eqref{eq: zitter}. 
After a sufficiently long time $t \gg T$, the center of mass moves linearly with $t$ parallel to $\bm{B}$ except for $\theta_0 = 0$. 
In particular, the long-time average of $X_{C, 3}(t)$ satisfies 
\begin{align}
	\lim_{\mathcal{T} \to \infty} 
	{1 \over \mathcal{T}}
	\int_0^{\mathcal{T}} dt \ X_{C, 3}(t) \ge 0
	, 
\end{align}
which manifests chiral transport consistent with the CME. 
When $\bm{S}$ is antiparallel to $\bm{B}$, i.e. $\theta_0 = \pi$, the wave packet occupies only the chiral band $\Psi_0$ and hence the linear dispersion gives a unidirectional motion $X_{C, 3} \propto t$ [red curve in Fig.~\ref{fig: single particle}]. 
When $\bm{S}$ is parallel to $\bm{B}$, i.e., $\theta_0 = 0$, the initial state is a superposition of the particle and hole states $\Psi_{\pm 1}$. 
Therefore, the time evolution $\bar{\mathcal{U}}_F$ induces an oscillation with the frequency given by their energy difference $2\omega$ [blue curve in Fig.~\ref{fig: single particle}]. 
One can see from the second last line of Eq.~\eqref{eq: zitter} that this oscillation results from the interference between the particle band $\Psi_1$ and the hole band $\Psi_{-1}$. 
This is analogous to the \textit{Zitterbewegung}, which was originally discussed for a Dirac particle in the absence of a magnetic field \cite{SchrodingerE30}, and later generalized to a relativistic particle under a magnetic field \cite{BermudezA07, RusinT10}. 

\section{\label{sec: classification}K-group correspondence in the classification of gapless Floquet states}
In this appendix, we derive the K-group correspondence between gapless Floquet states and static TIs/TSCs. The derivation proceeds in a manner similar to that for Floquet TIs/TSCs given in Refs.~\cite{RoyHarper17, MorimotoPoVishwanath17}. 
As mentioned in the main text, we construct from the Floquet operator $U(\bm{k})$ a Hermitian matrix 
\begin{align}
	H_U(\bm{k})=
\begin{pmatrix}
0 & U(\bm{k})\\
U^\dag(\bm{k}) & 0
\end{pmatrix}
 = \sig_- \otimes U^\dag(\bk) + \sig_+ \otimes U(\bk)
, \label{eq_HU2}
\end{align}
where $\otimes$ denotes the tensor product. 
This Hamiltonian has the following chiral symmetry (CS):
\begin{align}
\Gamma_2 H_U(\bm{k})\Gamma_2^{-1}=-H_U(\bm{k})
, 
\label{HU2_chiral2}
\end{align}
where 
\begin{align}
	\Gamma_2=
\begin{pmatrix}
\bm{1}_N & 0\\
0 & -\bm{1}_N
\end{pmatrix}
 = \sig_3 \otimes \bm{1}_N
. 
\end{align}
Therefore, the classification of class A (i.e., no symmetry) Floquet operators in $d$ dimensions is equivalent to that of class AIII TIs in $d$ dimensions. When the Floquet operator has the time-reversal symmetry (TRS) $\Theta U(\bm{k})\Theta^{-1}=U^\dag(-\bm{k})$ with $\epsilon_\Theta:=\Theta^2=\pm 1$, the Hamiltonian $H_U(\bm{k})$ has the following symmetries
\begin{align}
\Theta_1H_U(\bm{k})\Theta_1^{-1}&=H_U(-\bm{k})
,\label{HU2_T1} \\ 
\Theta_2H_U(\bm{k})\Theta_2^{-1}&=-H_U(-\bm{k})
, \label{HU2_T2}
\end{align}
where
\begin{align}
\Theta_1 &:=
\begin{pmatrix}
0 & \Theta\\
\Theta & 0
\end{pmatrix}
 = \sig_1 \otimes \Theta
 , \\
\Theta_2 &:= \epsilon_\Theta
\begin{pmatrix}
0 & -\Theta\\
\Theta & 0
\end{pmatrix} 
 = -i \epsilon_\Theta \sig_2 \otimes \Theta
. 
\end{align}
These symmetries are regarded as effective TRS and particle-hole symmetry (PHS) of $H_U(\bm{k})$. Since $\Theta_1^2=\epsilon_\Theta$ and $\Theta_2^2=-\epsilon_\Theta$, the classification of class AI (AII) Floquet operators in $d$ dimensions corresponds to that of class CI (DIII) TSCs in the same dimensions. Similarly, when $U(\bm{k})$ has the PHS $CU(\bm{k})C^{-1}=U(-\bm{k})$ with $\epsilon_C:=C^2=\pm 1$, we can define
\begin{align}
C_1 &:=
\begin{pmatrix}
C & 0\\
0 & C
\end{pmatrix} 
 = \sig_0 \otimes C
, \\ 
C_2 &:= \epsilon_C C_1\Gamma_2=\epsilon_C
\begin{pmatrix}
C & 0\\
0 & -C
\end{pmatrix} 
 = \epsilon_C \sig_3 \otimes C
,
\end{align}
so that
\begin{align}
C_1H_U(\bm{k})C_1^{-1} &= H_U(-\bm{k})
, \label{HU2_C1} \\
C_2H_U(\bm{k})C_2^{-1} &= -H_U(-\bm{k}),\label{HU2_C2}
\end{align}
with $C_1^2=C_2^2=\epsilon_C$. Therefore, the classification of class D (C) Floquet operators in $d$ dimensions corresponds to that of class BDI (CII) TSCs in the same dimensions.

Next, we consider the cases where the Floquet operator has the CS $\Gamma U(\bm{k})\Gamma^{-1}=U^\dag (\bm{k})$. Then, the Hamiltonian $H_U(\bm{k})$ satisfies
\begin{equation}
\Gamma_1 H_U(\bm{k})\Gamma_1^{-1}=-H_U(\bm{k})
, \label{HU2_chiral1}
\end{equation}
where $\Gamma_1$ is another CS of the Hamiltonian:
\begin{equation}
\Gamma_1:=
\begin{pmatrix}
0 & -\Gamma\\
\Gamma & 0
\end{pmatrix} 
 = -i \sig_2 \otimes \Gamma
.
\end{equation}
By combining the two chiral symmetries \eqref{HU2_chiral2} and \eqref{HU2_chiral1}, we have
\begin{equation}
\Gamma_1\Gamma_2H_U(\bm{k})(\Gamma_1\Gamma_2)^{-1}=H_U(\bm{k})
, 
\end{equation}
and therefore the Hamiltonian $H_U(\bm{k})$ can be block-diagonalized simultaneously with $\Gamma_1\Gamma_2$. 
We introduce $\gamma=1$ for $\Gamma^2=1$ and $\gamma=i$ for $\Gamma^2=-1$. 
Using a unitary matrix $V$ defined by 
\begin{align}
	V = {1 \over \sqrt{2}} \twmat{\gamma \Gamma}{-\gamma \Gamma}{\bm{1}_N}{\bm{1}_N}
	,
\end{align}
we obtain the block-diagonalized Hamiltonian and $\Gamma_1\Gamma_2$: 
\begin{align}
	V^\dag H_U(\bk)V &= \twmat{\gamma \Gamma U(\bk)}{0}{0}{-\gamma \Gamma U(\bk)}
	 = \sig_3 \otimes \gamma\Gamma U(\bk)
	, \\
	V^\dag \Gamma_1\Gamma_2 V &= \gamma^*\twmat{\bm{1}_N}{0}{0}{ -\bm{1}_N}
	 = \gamma^* \sig_3 \otimes \bm{1}_N
	. 
\end{align}
$\Gamma_{1,2}$ have off-diagonal forms in this basis:
\begin{align}
	V^\dag \Gamma_1 V &
	 = -i \gamma^* \sig_2 \otimes \bm{1}_N
	, \quad
	V^\dag \Gamma_2 V 
	 = - \sig_1 \otimes \bm{1}_N
	. 
\end{align}
Thus, the classification of $H_U(\bm{k})$ reduces to that of $H_D(\bm{k}) := \gamma \Gamma U(\bk)$. 
Note that the Hermiticity of $H_D(\bm{k})$ follows from the CS of $U(\bk)$. 
For example, if the Floquet operator has the CS only (i.e., class AIII), $H_D(\bm{k})$ is a Hamiltonian without symmetry. 
Therefore, the classification of class AIII Floquet operators in $d$ dimensions reduces to that of class A TIs in the same dimensions. 
For the remaining classes, where the Floquet operators have TRS, PHS, and CS, we identify the symmetry of $H_D(\bm{k})$ by using $\Gamma^2=\epsilon_\Theta\epsilon_C$. 
Here we assume that $\Theta$ and $C$ commute with each other: $[\Theta,C]=0$. 
Then, $\Theta_{1,2}$ and $C_{1,2}$ are expressed in the new basis as follows: 
\begin{align}
	V^\dag \Theta_1 V &= 
	\epsilon_\Theta (\sig_3 \mr{Re}\gamma - \sig_2\mr{Im}\gamma) \otimes C
	, \\
	V^\dag \Theta_2 V &= 
	-i (\sig_2\mr{Re}\gamma + \sig_3 \mr{Im}\gamma) \otimes C
	, \label{eq: fcme, 26}\\
	V^\dag C_1 V &= 
	\br{\sig_3 {1+\gamma^2 \over 2} + \sig_1{1-\gamma^2 \over 2}}
	\otimes C
	, \\
	V^\dag C_2 V &= 
	\br{\sig_0 {-1+\gamma^2 \over 2} - \sig_1{1+\gamma^2 \over 2}}
	\otimes C
	. \label{eq: fcme, 27}
\end{align}
When $\gamma^2 = 1$, we find from Eqs.~\eqref{eq: fcme, 26} and \eqref{eq: fcme, 27} that $\Theta_1$ and $C_1$ are symmetries of $H_D(\bm{k})$, while $\Theta_2$ and $C_2$ are not because $\Theta_2$ and $C_2$ do not preserve the eigenvalue of $\Gamma_1\Gamma_2$. 
The symmetries by $\Theta_1$ and $C_1$ [Eqs.~\eqref{HU2_T1} and \eqref{HU2_C1}] are equivalently expressed as 
\begin{equation}
C H_D(\bm{k})C^{-1}=H_D(-\bm{k})
\end{equation}
and this is the TRS of $H_D(\bm{k})$. 
Therefore, the classification of class BDI (CII) Floquet operators in $d$ dimensions is mapped to that of $d$-dimensional class AI (AII) TIs. 
When $\gamma^2 = -1$, we find from Eqs.~\eqref{eq: fcme, 26} and \eqref{eq: fcme, 27} that $\Theta_2$ and $C_2$ are symmetries of $H_D(\bm{k})$ while $\Theta_1$ and $C_1$ are not because $\Theta_1$ and $C_1$ do not preserve the eigenvalue of $\Gamma_1\Gamma_2$. 
The symmetries by $\Theta_2$ and $C_2$ [Eqs.~\eqref{HU2_T2} and \eqref{HU2_C2}] are equivalently expressed as \begin{equation}
C H_D(\bm{k})C^{-1}=-H_D(-\bm{k}),
\end{equation}
which is the PHS of $H_D(\bm{k})$. 
Thus we find that the classification of class DIII (CI) Floquet operators in $d$ dimensions is equivalent to that of class D (C) TSCs in the same dimensions.

We thus obtain the following results on the classification of Floquet operators. We denote the K-group of the Floquet operators by $K^{\mathrm{F}}_\mathbb{F}(s,d)$, where $d$ is the spatial dimension and $(\mathbb{F},s)$ represents the Altland-Zirnbauer symmetry class. To be concrete, $\mathbb{F}=\mathbb{C}$ with $s=0,1$ (mod $2$) corresponds to the complex class (A, AIII), and $\mathbb{F}=\mathbb{R}$ with $s=0,1,\cdots,7$ (mod $8$) corresponds to the real class (AI, BDI, D, DIII, AII, CII, C, CI). 
Similarly, we denote the K-group of static TIs/TSCs by $K_\mathbb{F}(s,d)$. Then, the analyses in this section indicate that
\begin{equation}
K^{\mathrm{F}}_\mathbb{F}(s,d)=K_\mathbb{F}(s-1,d).
\end{equation}
We list the resulting classification of Floquet operators in Table I in the main text. Using the K-group isomorphism $K_\mathbb{F}(s-1,d)=K_\mathbb{F}(s,d+1)$ \cite{Kitaev09, TeoKane10}, we obtain
\begin{equation}
K^{\mathrm{F}}_\mathbb{F}(s,d)=K_\mathbb{F}(s,d+1),
\end{equation}
which shows the equivalence between the class of gapless bulk states in Floquet systems and that of surface gapless states of static TIs/TSCs.

Finally, we note that the classification of gapless Floquet states in $d$ dimensions coincides with that of anomalous edge (or surface) states of Floquet TIs/TSCs given by unitary loops in $(d+1)$ dimensions \cite{RoyHarper17}. Indeed, it has been discussed that the gapless Floquet spectrum can be realized as the edge state of the anomalous Floquet TIs/TSCs \cite{Rudner13}. We may interpret our result to be a generalization of this correspondence to all the Altland-Zirnbauer classes. The correspondence between the gapless topological singularities and topologically nontrivial unitaries has been discussed in the context of anomalous Floquet TIs/TSCs \cite{NathanRudner15, RoyHarper17}.

\bibliography{FCME_ref.bib}

\begin{thebibliography}{95}
\expandafter\ifx\csname natexlab\endcsname\relax\def\natexlab#1{#1}\fi
\expandafter\ifx\csname bibnamefont\endcsname\relax
  \def\bibnamefont#1{#1}\fi
\expandafter\ifx\csname bibfnamefont\endcsname\relax
  \def\bibfnamefont#1{#1}\fi
\expandafter\ifx\csname citenamefont\endcsname\relax
  \def\citenamefont#1{#1}\fi
\expandafter\ifx\csname url\endcsname\relax
  \def\url#1{\texttt{#1}}\fi
\expandafter\ifx\csname urlprefix\endcsname\relax\def\urlprefix{URL }\fi
\providecommand{\bibinfo}[2]{#2}
\providecommand{\eprint}[2][]{\url{#2}}

\bibitem[{\citenamefont{Nielsen and
  Ninomiya}(1981{\natexlab{a}})}]{NielsenNinomiya1}
\bibinfo{author}{\bibfnamefont{H.}~\bibnamefont{Nielsen}} \bibnamefont{and}
  \bibinfo{author}{\bibfnamefont{M.}~\bibnamefont{Ninomiya}},
  \bibinfo{journal}{Nuclear Physics B} \textbf{\bibinfo{volume}{185}},
  \bibinfo{pages}{20 } (\bibinfo{year}{1981}{\natexlab{a}}), ISSN
  \bibinfo{issn}{0550-3213},
  \urlprefix\url{http://www.sciencedirect.com/science/article/pii/0550321381903618}.

\bibitem[{\citenamefont{Nielsen and
  Ninomiya}(1981{\natexlab{b}})}]{NielsenNinomiya2}
\bibinfo{author}{\bibfnamefont{H.}~\bibnamefont{Nielsen}} \bibnamefont{and}
  \bibinfo{author}{\bibfnamefont{M.}~\bibnamefont{Ninomiya}},
  \bibinfo{journal}{Nuclear Physics B} \textbf{\bibinfo{volume}{193}},
  \bibinfo{pages}{173 } (\bibinfo{year}{1981}{\natexlab{b}}), ISSN
  \bibinfo{issn}{0550-3213},
  \urlprefix\url{http://www.sciencedirect.com/science/article/pii/0550321381905241}.

\bibitem[{\citenamefont{Murakami}(2007)}]{Murakami07}
\bibinfo{author}{\bibfnamefont{S.}~\bibnamefont{Murakami}},
  \bibinfo{journal}{New Journal of Physics} \textbf{\bibinfo{volume}{9}},
  \bibinfo{pages}{356} (\bibinfo{year}{2007}),
  \urlprefix\url{http://stacks.iop.org/1367-2630/9/i=9/a=356}.

\bibitem[{\citenamefont{Wan et~al.}(2011)\citenamefont{Wan, Turner, Vishwanath,
  and Savrasov}}]{Wan11}
\bibinfo{author}{\bibfnamefont{X.}~\bibnamefont{Wan}},
  \bibinfo{author}{\bibfnamefont{A.~M.} \bibnamefont{Turner}},
  \bibinfo{author}{\bibfnamefont{A.}~\bibnamefont{Vishwanath}},
  \bibnamefont{and} \bibinfo{author}{\bibfnamefont{S.~Y.}
  \bibnamefont{Savrasov}}, \bibinfo{journal}{Phys. Rev. B}
  \textbf{\bibinfo{volume}{83}}, \bibinfo{pages}{205101}
  (\bibinfo{year}{2011}),
  \urlprefix\url{https://link.aps.org/doi/10.1103/PhysRevB.83.205101}.

\bibitem[{\citenamefont{Nielsen and Ninomiya}(1983)}]{NielsenNinomiya3}
\bibinfo{author}{\bibfnamefont{H.}~\bibnamefont{Nielsen}} \bibnamefont{and}
  \bibinfo{author}{\bibfnamefont{M.}~\bibnamefont{Ninomiya}},
  \bibinfo{journal}{Physics Letters B} \textbf{\bibinfo{volume}{130}},
  \bibinfo{pages}{389 } (\bibinfo{year}{1983}), ISSN \bibinfo{issn}{0370-2693},
  \urlprefix\url{http://www.sciencedirect.com/science/article/pii/0370269383915290}.

\bibitem[{\citenamefont{Zyuzin and Burkov}(2012)}]{Zyuzin12}
\bibinfo{author}{\bibfnamefont{A.~A.} \bibnamefont{Zyuzin}} \bibnamefont{and}
  \bibinfo{author}{\bibfnamefont{A.~A.} \bibnamefont{Burkov}},
  \bibinfo{journal}{Phys. Rev. B} \textbf{\bibinfo{volume}{86}},
  \bibinfo{pages}{115133} (\bibinfo{year}{2012}),
  \urlprefix\url{https://link.aps.org/doi/10.1103/PhysRevB.86.115133}.

\bibitem[{\citenamefont{Son and Yamamoto}(2012)}]{Son12}
\bibinfo{author}{\bibfnamefont{D.~T.} \bibnamefont{Son}} \bibnamefont{and}
  \bibinfo{author}{\bibfnamefont{N.}~\bibnamefont{Yamamoto}},
  \bibinfo{journal}{Phys. Rev. Lett.} \textbf{\bibinfo{volume}{109}},
  \bibinfo{pages}{181602} (\bibinfo{year}{2012}),
  \urlprefix\url{https://link.aps.org/doi/10.1103/PhysRevLett.109.181602}.

\bibitem[{\citenamefont{Burkov}(2014)}]{Burkov14}
\bibinfo{author}{\bibfnamefont{A.~A.} \bibnamefont{Burkov}},
  \bibinfo{journal}{Phys. Rev. Lett.} \textbf{\bibinfo{volume}{113}},
  \bibinfo{pages}{247203} (\bibinfo{year}{2014}),
  \urlprefix\url{https://link.aps.org/doi/10.1103/PhysRevLett.113.247203}.

\bibitem[{\citenamefont{Lv et~al.}(2015{\natexlab{a}})\citenamefont{Lv, Weng,
  Fu, Wang, Miao, Ma, Richard, Huang, Zhao, Chen et~al.}}]{Lv15_1}
\bibinfo{author}{\bibfnamefont{B.~Q.} \bibnamefont{Lv}},
  \bibinfo{author}{\bibfnamefont{H.~M.} \bibnamefont{Weng}},
  \bibinfo{author}{\bibfnamefont{B.~B.} \bibnamefont{Fu}},
  \bibinfo{author}{\bibfnamefont{X.~P.} \bibnamefont{Wang}},
  \bibinfo{author}{\bibfnamefont{H.}~\bibnamefont{Miao}},
  \bibinfo{author}{\bibfnamefont{J.}~\bibnamefont{Ma}},
  \bibinfo{author}{\bibfnamefont{P.}~\bibnamefont{Richard}},
  \bibinfo{author}{\bibfnamefont{X.~C.} \bibnamefont{Huang}},
  \bibinfo{author}{\bibfnamefont{L.~X.} \bibnamefont{Zhao}},
  \bibinfo{author}{\bibfnamefont{G.~F.} \bibnamefont{Chen}},
  \bibnamefont{et~al.}, \bibinfo{journal}{Phys. Rev. X}
  \textbf{\bibinfo{volume}{5}}, \bibinfo{pages}{031013}
  (\bibinfo{year}{2015}{\natexlab{a}}),
  \urlprefix\url{https://link.aps.org/doi/10.1103/PhysRevX.5.031013}.

\bibitem[{\citenamefont{Xu et~al.}(2015)\citenamefont{Xu, Belopolski, Alidoust,
  Neupane, Bian, Zhang, Sankar, Chang, Yuan, Lee et~al.}}]{SYXu15}
\bibinfo{author}{\bibfnamefont{S.-Y.} \bibnamefont{Xu}},
  \bibinfo{author}{\bibfnamefont{I.}~\bibnamefont{Belopolski}},
  \bibinfo{author}{\bibfnamefont{N.}~\bibnamefont{Alidoust}},
  \bibinfo{author}{\bibfnamefont{M.}~\bibnamefont{Neupane}},
  \bibinfo{author}{\bibfnamefont{G.}~\bibnamefont{Bian}},
  \bibinfo{author}{\bibfnamefont{C.}~\bibnamefont{Zhang}},
  \bibinfo{author}{\bibfnamefont{R.}~\bibnamefont{Sankar}},
  \bibinfo{author}{\bibfnamefont{G.}~\bibnamefont{Chang}},
  \bibinfo{author}{\bibfnamefont{Z.}~\bibnamefont{Yuan}},
  \bibinfo{author}{\bibfnamefont{C.-C.} \bibnamefont{Lee}},
  \bibnamefont{et~al.}, \bibinfo{journal}{Science}
  \textbf{\bibinfo{volume}{349}}, \bibinfo{pages}{613} (\bibinfo{year}{2015}),
  ISSN \bibinfo{issn}{0036-8075},
  \urlprefix\url{http://science.sciencemag.org/content/349/6248/613}.

\bibitem[{\citenamefont{Lv et~al.}(2015{\natexlab{b}})\citenamefont{Lv, Xu,
  Weng, Ma, Richard, Huang, Zhao, Chen, Matt, Bisti et~al.}}]{Lv15_2}
\bibinfo{author}{\bibfnamefont{B.~Q.} \bibnamefont{Lv}},
  \bibinfo{author}{\bibfnamefont{N.}~\bibnamefont{Xu}},
  \bibinfo{author}{\bibfnamefont{H.~M.} \bibnamefont{Weng}},
  \bibinfo{author}{\bibfnamefont{J.~Z.} \bibnamefont{Ma}},
  \bibinfo{author}{\bibfnamefont{P.}~\bibnamefont{Richard}},
  \bibinfo{author}{\bibfnamefont{X.~C.} \bibnamefont{Huang}},
  \bibinfo{author}{\bibfnamefont{L.~X.} \bibnamefont{Zhao}},
  \bibinfo{author}{\bibfnamefont{G.~F.} \bibnamefont{Chen}},
  \bibinfo{author}{\bibfnamefont{C.~E.} \bibnamefont{Matt}},
  \bibinfo{author}{\bibfnamefont{F.}~\bibnamefont{Bisti}},
  \bibnamefont{et~al.}, \bibinfo{journal}{Nat. Phys.}
  \textbf{\bibinfo{volume}{11}}, \bibinfo{pages}{724}
  (\bibinfo{year}{2015}{\natexlab{b}}),
  \urlprefix\url{https://www.nature.com/articles/nphys3426}.

\bibitem[{\citenamefont{Yang et~al.}(2015)\citenamefont{Yang, Liu, Sun, Peng,
  Yang, Zhang, Zhou, Zhang, Guo, Rahn et~al.}}]{LXYang15}
\bibinfo{author}{\bibfnamefont{L.~X.} \bibnamefont{Yang}},
  \bibinfo{author}{\bibfnamefont{Z.~K.} \bibnamefont{Liu}},
  \bibinfo{author}{\bibfnamefont{Y.}~\bibnamefont{Sun}},
  \bibinfo{author}{\bibfnamefont{H.}~\bibnamefont{Peng}},
  \bibinfo{author}{\bibfnamefont{H.~F.} \bibnamefont{Yang}},
  \bibinfo{author}{\bibfnamefont{T.}~\bibnamefont{Zhang}},
  \bibinfo{author}{\bibfnamefont{B.}~\bibnamefont{Zhou}},
  \bibinfo{author}{\bibfnamefont{Y.}~\bibnamefont{Zhang}},
  \bibinfo{author}{\bibfnamefont{Y.~F.} \bibnamefont{Guo}},
  \bibinfo{author}{\bibfnamefont{M.}~\bibnamefont{Rahn}}, \bibnamefont{et~al.},
  \bibinfo{journal}{Nat. Phys.} \textbf{\bibinfo{volume}{11}},
  \bibinfo{pages}{728} (\bibinfo{year}{2015}),
  \urlprefix\url{https://www.nature.com/articles/nphys3425}.

\bibitem[{\citenamefont{Xu et~al.}(2016{\natexlab{a}})\citenamefont{Xu,
  Belopolski, Sanchez, Neupane, Chang, Yaji, Yuan, Zhang, Kuroda, Bian
  et~al.}}]{SYXu16}
\bibinfo{author}{\bibfnamefont{S.-Y.} \bibnamefont{Xu}},
  \bibinfo{author}{\bibfnamefont{I.}~\bibnamefont{Belopolski}},
  \bibinfo{author}{\bibfnamefont{D.~S.} \bibnamefont{Sanchez}},
  \bibinfo{author}{\bibfnamefont{M.}~\bibnamefont{Neupane}},
  \bibinfo{author}{\bibfnamefont{G.}~\bibnamefont{Chang}},
  \bibinfo{author}{\bibfnamefont{K.}~\bibnamefont{Yaji}},
  \bibinfo{author}{\bibfnamefont{Z.}~\bibnamefont{Yuan}},
  \bibinfo{author}{\bibfnamefont{C.}~\bibnamefont{Zhang}},
  \bibinfo{author}{\bibfnamefont{K.}~\bibnamefont{Kuroda}},
  \bibinfo{author}{\bibfnamefont{G.}~\bibnamefont{Bian}}, \bibnamefont{et~al.},
  \bibinfo{journal}{Phys. Rev. Lett.} \textbf{\bibinfo{volume}{116}},
  \bibinfo{pages}{096801} (\bibinfo{year}{2016}{\natexlab{a}}),
  \urlprefix\url{https://link.aps.org/doi/10.1103/PhysRevLett.116.096801}.

\bibitem[{\citenamefont{Xu et~al.}(2016{\natexlab{b}})\citenamefont{Xu, Weng,
  Lv, Matt, Park, Bisti, Strocov, Gawryluk, Pomjakushina, Conder
  et~al.}}]{NXu16}
\bibinfo{author}{\bibfnamefont{N.}~\bibnamefont{Xu}},
  \bibinfo{author}{\bibfnamefont{H.~M.} \bibnamefont{Weng}},
  \bibinfo{author}{\bibfnamefont{B.~Q.} \bibnamefont{Lv}},
  \bibinfo{author}{\bibfnamefont{C.~E.} \bibnamefont{Matt}},
  \bibinfo{author}{\bibfnamefont{J.}~\bibnamefont{Park}},
  \bibinfo{author}{\bibfnamefont{F.}~\bibnamefont{Bisti}},
  \bibinfo{author}{\bibfnamefont{V.~N.} \bibnamefont{Strocov}},
  \bibinfo{author}{\bibfnamefont{D.}~\bibnamefont{Gawryluk}},
  \bibinfo{author}{\bibfnamefont{E.}~\bibnamefont{Pomjakushina}},
  \bibinfo{author}{\bibfnamefont{K.}~\bibnamefont{Conder}},
  \bibnamefont{et~al.}, \bibinfo{journal}{Nat. Commun.}
  \textbf{\bibinfo{volume}{7}}, \bibinfo{pages}{11006}
  (\bibinfo{year}{2016}{\natexlab{b}}),
  \urlprefix\url{https://www.nature.com/articles/ncomms11006}.

\bibitem[{\citenamefont{Huang et~al.}(2015{\natexlab{a}})\citenamefont{Huang,
  Xu, Belopolski, Lee, Chang, Wang, Alidoust, Bian, Neupane, Zhang
  et~al.}}]{SMHuang15}
\bibinfo{author}{\bibfnamefont{S.-M.} \bibnamefont{Huang}},
  \bibinfo{author}{\bibfnamefont{S.-Y.} \bibnamefont{Xu}},
  \bibinfo{author}{\bibfnamefont{I.}~\bibnamefont{Belopolski}},
  \bibinfo{author}{\bibfnamefont{C.-C.} \bibnamefont{Lee}},
  \bibinfo{author}{\bibfnamefont{G.}~\bibnamefont{Chang}},
  \bibinfo{author}{\bibfnamefont{B.}~\bibnamefont{Wang}},
  \bibinfo{author}{\bibfnamefont{N.}~\bibnamefont{Alidoust}},
  \bibinfo{author}{\bibfnamefont{G.}~\bibnamefont{Bian}},
  \bibinfo{author}{\bibfnamefont{M.}~\bibnamefont{Neupane}},
  \bibinfo{author}{\bibfnamefont{C.}~\bibnamefont{Zhang}},
  \bibnamefont{et~al.}, \bibinfo{journal}{Nat. Commun.}
  \textbf{\bibinfo{volume}{6}}, \bibinfo{pages}{7373}
  (\bibinfo{year}{2015}{\natexlab{a}}),
  \urlprefix\url{https://www.nature.com/articles/ncomms8373}.

\bibitem[{\citenamefont{Huang et~al.}(2015{\natexlab{b}})\citenamefont{Huang,
  Zhao, Long, Wang, Chen, Yang, Liang, Xue, Weng, Fang et~al.}}]{XHuang15}
\bibinfo{author}{\bibfnamefont{X.}~\bibnamefont{Huang}},
  \bibinfo{author}{\bibfnamefont{L.}~\bibnamefont{Zhao}},
  \bibinfo{author}{\bibfnamefont{Y.}~\bibnamefont{Long}},
  \bibinfo{author}{\bibfnamefont{P.}~\bibnamefont{Wang}},
  \bibinfo{author}{\bibfnamefont{D.}~\bibnamefont{Chen}},
  \bibinfo{author}{\bibfnamefont{Z.}~\bibnamefont{Yang}},
  \bibinfo{author}{\bibfnamefont{H.}~\bibnamefont{Liang}},
  \bibinfo{author}{\bibfnamefont{M.}~\bibnamefont{Xue}},
  \bibinfo{author}{\bibfnamefont{H.}~\bibnamefont{Weng}},
  \bibinfo{author}{\bibfnamefont{Z.}~\bibnamefont{Fang}}, \bibnamefont{et~al.},
  \bibinfo{journal}{Phys. Rev. X} \textbf{\bibinfo{volume}{5}},
  \bibinfo{pages}{031023} (\bibinfo{year}{2015}{\natexlab{b}}),
  \urlprefix\url{https://link.aps.org/doi/10.1103/PhysRevX.5.031023}.

\bibitem[{\citenamefont{Arnold et~al.}(2016)\citenamefont{Arnold, Shekhar, Wu,
  Sun, dos Reis, Kumar, Naumann, Ajeesh, Schmidt, Grushin et~al.}}]{Arnold16}
\bibinfo{author}{\bibfnamefont{F.}~\bibnamefont{Arnold}},
  \bibinfo{author}{\bibfnamefont{C.}~\bibnamefont{Shekhar}},
  \bibinfo{author}{\bibfnamefont{S.-C.} \bibnamefont{Wu}},
  \bibinfo{author}{\bibfnamefont{Y.}~\bibnamefont{Sun}},
  \bibinfo{author}{\bibfnamefont{R.~D.} \bibnamefont{dos Reis}},
  \bibinfo{author}{\bibfnamefont{N.}~\bibnamefont{Kumar}},
  \bibinfo{author}{\bibfnamefont{M.}~\bibnamefont{Naumann}},
  \bibinfo{author}{\bibfnamefont{M.~O.} \bibnamefont{Ajeesh}},
  \bibinfo{author}{\bibfnamefont{M.}~\bibnamefont{Schmidt}},
  \bibinfo{author}{\bibfnamefont{A.~G.} \bibnamefont{Grushin}},
  \bibnamefont{et~al.}, \bibinfo{journal}{Nat. Commun.}
  \textbf{\bibinfo{volume}{7}}, \bibinfo{pages}{11615} (\bibinfo{year}{2016}),
  \urlprefix\url{https://www.nature.com/articles/ncomms11615}.

\bibitem[{\citenamefont{Wang et~al.}(2016{\natexlab{a}})\citenamefont{Wang,
  Zheng, Shen, Lu, Fang, Sheng, Zhou, Yang, Li, Feng et~al.}}]{Zhen16}
\bibinfo{author}{\bibfnamefont{Z.}~\bibnamefont{Wang}},
  \bibinfo{author}{\bibfnamefont{Y.}~\bibnamefont{Zheng}},
  \bibinfo{author}{\bibfnamefont{Z.}~\bibnamefont{Shen}},
  \bibinfo{author}{\bibfnamefont{Y.}~\bibnamefont{Lu}},
  \bibinfo{author}{\bibfnamefont{H.}~\bibnamefont{Fang}},
  \bibinfo{author}{\bibfnamefont{F.}~\bibnamefont{Sheng}},
  \bibinfo{author}{\bibfnamefont{Y.}~\bibnamefont{Zhou}},
  \bibinfo{author}{\bibfnamefont{X.}~\bibnamefont{Yang}},
  \bibinfo{author}{\bibfnamefont{Y.}~\bibnamefont{Li}},
  \bibinfo{author}{\bibfnamefont{C.}~\bibnamefont{Feng}}, \bibnamefont{et~al.},
  \bibinfo{journal}{Phys. Rev. B} \textbf{\bibinfo{volume}{93}},
  \bibinfo{pages}{121112} (\bibinfo{year}{2016}{\natexlab{a}}),
  \urlprefix\url{https://link.aps.org/doi/10.1103/PhysRevB.93.121112}.

\bibitem[{\citenamefont{Zhang et~al.}(2016{\natexlab{a}})\citenamefont{Zhang,
  Xu, Belopolski, Yuan, Lin, Tong, Bian, Alidoust, Lee, Huang
  et~al.}}]{ZhangHasan16}
\bibinfo{author}{\bibfnamefont{C.-L.} \bibnamefont{Zhang}},
  \bibinfo{author}{\bibfnamefont{S.-Y.} \bibnamefont{Xu}},
  \bibinfo{author}{\bibfnamefont{I.}~\bibnamefont{Belopolski}},
  \bibinfo{author}{\bibfnamefont{Z.}~\bibnamefont{Yuan}},
  \bibinfo{author}{\bibfnamefont{Z.}~\bibnamefont{Lin}},
  \bibinfo{author}{\bibfnamefont{B.}~\bibnamefont{Tong}},
  \bibinfo{author}{\bibfnamefont{G.}~\bibnamefont{Bian}},
  \bibinfo{author}{\bibfnamefont{N.}~\bibnamefont{Alidoust}},
  \bibinfo{author}{\bibfnamefont{C.-C.} \bibnamefont{Lee}},
  \bibinfo{author}{\bibfnamefont{S.-M.} \bibnamefont{Huang}},
  \bibnamefont{et~al.}, \bibinfo{journal}{Nature Communications}
  \textbf{\bibinfo{volume}{7}}, \bibinfo{pages}{10735}
  (\bibinfo{year}{2016}{\natexlab{a}}),
  \urlprefix\url{http://dx.doi.org/10.1038/ncomms10735
  http://10.0.4.14/ncomms10735
  https://www.nature.com/articles/ncomms10735{\#}supplementary-information}.

\bibitem[{\citenamefont{Fukushima et~al.}(2008)\citenamefont{Fukushima,
  Kharzeev, and Warringa}}]{Fukushima08}
\bibinfo{author}{\bibfnamefont{K.}~\bibnamefont{Fukushima}},
  \bibinfo{author}{\bibfnamefont{D.~E.} \bibnamefont{Kharzeev}},
  \bibnamefont{and} \bibinfo{author}{\bibfnamefont{H.~J.}
  \bibnamefont{Warringa}}, \bibinfo{journal}{Phys. Rev. D}
  \textbf{\bibinfo{volume}{78}}, \bibinfo{pages}{074033}
  (\bibinfo{year}{2008}),
  \urlprefix\url{https://link.aps.org/doi/10.1103/PhysRevD.78.074033}.

\bibitem[{\citenamefont{Vazifeh and Franz}(2013)}]{Vazifeh13}
\bibinfo{author}{\bibfnamefont{M.~M.} \bibnamefont{Vazifeh}} \bibnamefont{and}
  \bibinfo{author}{\bibfnamefont{M.}~\bibnamefont{Franz}},
  \bibinfo{journal}{Phys. Rev. Lett.} \textbf{\bibinfo{volume}{111}},
  \bibinfo{pages}{027201} (\bibinfo{year}{2013}),
  \urlprefix\url{https://link.aps.org/doi/10.1103/PhysRevLett.111.027201}.

\bibitem[{\citenamefont{Goswami et~al.}(2015)\citenamefont{Goswami, Sharma, and
  Tewari}}]{Goswami15}
\bibinfo{author}{\bibfnamefont{P.}~\bibnamefont{Goswami}},
  \bibinfo{author}{\bibfnamefont{G.}~\bibnamefont{Sharma}}, \bibnamefont{and}
  \bibinfo{author}{\bibfnamefont{S.}~\bibnamefont{Tewari}},
  \bibinfo{journal}{Phys. Rev. B} \textbf{\bibinfo{volume}{92}},
  \bibinfo{pages}{161110} (\bibinfo{year}{2015}),
  \urlprefix\url{https://link.aps.org/doi/10.1103/PhysRevB.92.161110}.

\bibitem[{\citenamefont{Taguchi et~al.}(2016)\citenamefont{Taguchi, Imaeda,
  Sato, and Tanaka}}]{Taguchi16}
\bibinfo{author}{\bibfnamefont{K.}~\bibnamefont{Taguchi}},
  \bibinfo{author}{\bibfnamefont{T.}~\bibnamefont{Imaeda}},
  \bibinfo{author}{\bibfnamefont{M.}~\bibnamefont{Sato}}, \bibnamefont{and}
  \bibinfo{author}{\bibfnamefont{Y.}~\bibnamefont{Tanaka}},
  \bibinfo{journal}{Phys. Rev. B} \textbf{\bibinfo{volume}{93}},
  \bibinfo{pages}{201202} (\bibinfo{year}{2016}),
  \urlprefix\url{https://link.aps.org/doi/10.1103/PhysRevB.93.201202}.

\bibitem[{\citenamefont{Sumiyoshi and Fujimoto}(2016)}]{Sumiyoshi16}
\bibinfo{author}{\bibfnamefont{H.}~\bibnamefont{Sumiyoshi}} \bibnamefont{and}
  \bibinfo{author}{\bibfnamefont{S.}~\bibnamefont{Fujimoto}},
  \bibinfo{journal}{Phys. Rev. Lett.} \textbf{\bibinfo{volume}{116}},
  \bibinfo{pages}{166601} (\bibinfo{year}{2016}),
  \urlprefix\url{https://link.aps.org/doi/10.1103/PhysRevLett.116.166601}.

\bibitem[{\citenamefont{Ibe and Sumiyoshi}(2017)}]{Ibe17}
\bibinfo{author}{\bibfnamefont{Y.}~\bibnamefont{Ibe}} \bibnamefont{and}
  \bibinfo{author}{\bibfnamefont{H.}~\bibnamefont{Sumiyoshi}},
  \bibinfo{journal}{Journal of the Physical Society of Japan}
  \textbf{\bibinfo{volume}{86}}, \bibinfo{pages}{054707}
  (\bibinfo{year}{2017}),
  \urlprefix\url{https://doi.org/10.7566/JPSJ.86.054707}.

\bibitem[{\citenamefont{Cortijo et~al.}(2016)\citenamefont{Cortijo, Kharzeev,
  Landsteiner, and Vozmediano}}]{Cortijo16}
\bibinfo{author}{\bibfnamefont{A.}~\bibnamefont{Cortijo}},
  \bibinfo{author}{\bibfnamefont{D.}~\bibnamefont{Kharzeev}},
  \bibinfo{author}{\bibfnamefont{K.}~\bibnamefont{Landsteiner}},
  \bibnamefont{and} \bibinfo{author}{\bibfnamefont{M.~A.~H.}
  \bibnamefont{Vozmediano}}, \bibinfo{journal}{Phys. Rev. B}
  \textbf{\bibinfo{volume}{94}}, \bibinfo{pages}{241405}
  (\bibinfo{year}{2016}),
  \urlprefix\url{https://link.aps.org/doi/10.1103/PhysRevB.94.241405}.

\bibitem[{\citenamefont{Pikulin et~al.}(2016)\citenamefont{Pikulin, Chen, and
  Franz}}]{Pikulin16}
\bibinfo{author}{\bibfnamefont{D.~I.} \bibnamefont{Pikulin}},
  \bibinfo{author}{\bibfnamefont{A.}~\bibnamefont{Chen}}, \bibnamefont{and}
  \bibinfo{author}{\bibfnamefont{M.}~\bibnamefont{Franz}},
  \bibinfo{journal}{Phys. Rev. X} \textbf{\bibinfo{volume}{6}},
  \bibinfo{pages}{041021} (\bibinfo{year}{2016}),
  \urlprefix\url{https://link.aps.org/doi/10.1103/PhysRevX.6.041021}.

\bibitem[{\citenamefont{Meng and Budich}(2018)}]{MengT18}
\bibinfo{author}{\bibfnamefont{T.}~\bibnamefont{Meng}} \bibnamefont{and}
  \bibinfo{author}{\bibfnamefont{J.~C.} \bibnamefont{Budich}},
  \bibinfo{journal}{arXiv preprint arXiv:1804.05078}  (\bibinfo{year}{2018}).

\bibitem[{\citenamefont{Chernodub and Cortijo}(2019)}]{ChernodubM19}
\bibinfo{author}{\bibfnamefont{M.~N.} \bibnamefont{Chernodub}}
  \bibnamefont{and} \bibinfo{author}{\bibfnamefont{A.}~\bibnamefont{Cortijo}},
  \bibinfo{journal}{arXiv preprint arXiv:1901.06167}  (\bibinfo{year}{2019}),
  \eprint{1901.06167}, \urlprefix\url{https://arxiv.org/abs/1901.06167}.

\bibitem[{\citenamefont{Eckardt}(2017)}]{Eckardt17}
\bibinfo{author}{\bibfnamefont{A.}~\bibnamefont{Eckardt}},
  \bibinfo{journal}{Rev. Mod. Phys.} \textbf{\bibinfo{volume}{89}},
  \bibinfo{pages}{011004} (\bibinfo{year}{2017}),
  \urlprefix\url{https://link.aps.org/doi/10.1103/RevModPhys.89.011004}.

\bibitem[{\citenamefont{Kitagawa
  et~al.}(2010{\natexlab{a}})\citenamefont{Kitagawa, Berg, Rudner, and
  Demler}}]{Kitagawa10}
\bibinfo{author}{\bibfnamefont{T.}~\bibnamefont{Kitagawa}},
  \bibinfo{author}{\bibfnamefont{E.}~\bibnamefont{Berg}},
  \bibinfo{author}{\bibfnamefont{M.}~\bibnamefont{Rudner}}, \bibnamefont{and}
  \bibinfo{author}{\bibfnamefont{E.}~\bibnamefont{Demler}},
  \bibinfo{journal}{Phys. Rev. B} \textbf{\bibinfo{volume}{82}},
  \bibinfo{pages}{235114} (\bibinfo{year}{2010}{\natexlab{a}}),
  \urlprefix\url{https://link.aps.org/doi/10.1103/PhysRevB.82.235114}.

\bibitem[{\citenamefont{Rudner et~al.}(2013)\citenamefont{Rudner, Lindner,
  Berg, and Levin}}]{Rudner13}
\bibinfo{author}{\bibfnamefont{M.~S.} \bibnamefont{Rudner}},
  \bibinfo{author}{\bibfnamefont{N.~H.} \bibnamefont{Lindner}},
  \bibinfo{author}{\bibfnamefont{E.}~\bibnamefont{Berg}}, \bibnamefont{and}
  \bibinfo{author}{\bibfnamefont{M.}~\bibnamefont{Levin}},
  \bibinfo{journal}{Phys. Rev. X} \textbf{\bibinfo{volume}{3}},
  \bibinfo{pages}{031005} (\bibinfo{year}{2013}),
  \urlprefix\url{https://link.aps.org/doi/10.1103/PhysRevX.3.031005}.

\bibitem[{\citenamefont{Else et~al.}(2016)\citenamefont{Else, Bauer, and
  Nayak}}]{Else16}
\bibinfo{author}{\bibfnamefont{D.~V.} \bibnamefont{Else}},
  \bibinfo{author}{\bibfnamefont{B.}~\bibnamefont{Bauer}}, \bibnamefont{and}
  \bibinfo{author}{\bibfnamefont{C.}~\bibnamefont{Nayak}},
  \bibinfo{journal}{Phys. Rev. Lett.} \textbf{\bibinfo{volume}{117}},
  \bibinfo{pages}{090402} (\bibinfo{year}{2016}),
  \urlprefix\url{https://link.aps.org/doi/10.1103/PhysRevLett.117.090402}.

\bibitem[{\citenamefont{Yao et~al.}(2017)\citenamefont{Yao, Potter, Potirniche,
  and Vishwanath}}]{Yao17}
\bibinfo{author}{\bibfnamefont{N.~Y.} \bibnamefont{Yao}},
  \bibinfo{author}{\bibfnamefont{A.~C.} \bibnamefont{Potter}},
  \bibinfo{author}{\bibfnamefont{I.-D.} \bibnamefont{Potirniche}},
  \bibnamefont{and}
  \bibinfo{author}{\bibfnamefont{A.}~\bibnamefont{Vishwanath}},
  \bibinfo{journal}{Phys. Rev. Lett.} \textbf{\bibinfo{volume}{118}},
  \bibinfo{pages}{030401} (\bibinfo{year}{2017}),
  \urlprefix\url{https://link.aps.org/doi/10.1103/PhysRevLett.118.030401}.

\bibitem[{\citenamefont{Thouless}(1983)}]{Thouless83}
\bibinfo{author}{\bibfnamefont{D.~J.} \bibnamefont{Thouless}},
  \bibinfo{journal}{Phys. Rev. B} \textbf{\bibinfo{volume}{27}},
  \bibinfo{pages}{6083} (\bibinfo{year}{1983}),
  \urlprefix\url{https://link.aps.org/doi/10.1103/PhysRevB.27.6083}.

\bibitem[{\citenamefont{Nakajima et~al.}(2016)\citenamefont{Nakajima, Tomita,
  Taie, Ichinose, Ozawa, Wang, Troyer, and Takahashi}}]{Nakajima16}
\bibinfo{author}{\bibfnamefont{S.}~\bibnamefont{Nakajima}},
  \bibinfo{author}{\bibfnamefont{T.}~\bibnamefont{Tomita}},
  \bibinfo{author}{\bibfnamefont{S.}~\bibnamefont{Taie}},
  \bibinfo{author}{\bibfnamefont{T.}~\bibnamefont{Ichinose}},
  \bibinfo{author}{\bibfnamefont{H.}~\bibnamefont{Ozawa}},
  \bibinfo{author}{\bibfnamefont{L.}~\bibnamefont{Wang}},
  \bibinfo{author}{\bibfnamefont{M.}~\bibnamefont{Troyer}}, \bibnamefont{and}
  \bibinfo{author}{\bibfnamefont{Y.}~\bibnamefont{Takahashi}},
  \bibinfo{journal}{Nat. Phys.} \textbf{\bibinfo{volume}{12}},
  \bibinfo{pages}{296} (\bibinfo{year}{2016}),
  \urlprefix\url{http://www.nature.com/articles/nphys3622}.

\bibitem[{\citenamefont{Lohse et~al.}(2016)\citenamefont{Lohse, Schweizer,
  Zilberberg, Aidelsburger, and Bloch}}]{Lohse16}
\bibinfo{author}{\bibfnamefont{M.}~\bibnamefont{Lohse}},
  \bibinfo{author}{\bibfnamefont{C.}~\bibnamefont{Schweizer}},
  \bibinfo{author}{\bibfnamefont{O.}~\bibnamefont{Zilberberg}},
  \bibinfo{author}{\bibfnamefont{M.}~\bibnamefont{Aidelsburger}},
  \bibnamefont{and} \bibinfo{author}{\bibfnamefont{I.}~\bibnamefont{Bloch}},
  \bibinfo{journal}{Nat. Phys.} \textbf{\bibinfo{volume}{12}},
  \bibinfo{pages}{350} (\bibinfo{year}{2016}),
  \urlprefix\url{http://www.nature.com/articles/nphys3584}.

\bibitem[{\citenamefont{Qi and Zhang}(2011)}]{QiZhang11}
\bibinfo{author}{\bibfnamefont{X.-L.} \bibnamefont{Qi}} \bibnamefont{and}
  \bibinfo{author}{\bibfnamefont{S.-C.} \bibnamefont{Zhang}},
  \bibinfo{journal}{Rev. Mod. Phys.} \textbf{\bibinfo{volume}{83}},
  \bibinfo{pages}{1057} (\bibinfo{year}{2011}),
  \urlprefix\url{https://link.aps.org/doi/10.1103/RevModPhys.83.1057}.

\bibitem[{\citenamefont{Ryu and Zhang}(2012)}]{RyuZhang12}
\bibinfo{author}{\bibfnamefont{S.}~\bibnamefont{Ryu}} \bibnamefont{and}
  \bibinfo{author}{\bibfnamefont{S.-C.} \bibnamefont{Zhang}},
  \bibinfo{journal}{Phys. Rev. B} \textbf{\bibinfo{volume}{85}},
  \bibinfo{pages}{245132} (\bibinfo{year}{2012}),
  \urlprefix\url{https://link.aps.org/doi/10.1103/PhysRevB.85.245132}.

\bibitem[{\citenamefont{Sule et~al.}(2013)\citenamefont{Sule, Chen, and
  Ryu}}]{Sule13}
\bibinfo{author}{\bibfnamefont{O.~M.} \bibnamefont{Sule}},
  \bibinfo{author}{\bibfnamefont{X.}~\bibnamefont{Chen}}, \bibnamefont{and}
  \bibinfo{author}{\bibfnamefont{S.}~\bibnamefont{Ryu}},
  \bibinfo{journal}{Phys. Rev. B} \textbf{\bibinfo{volume}{88}},
  \bibinfo{pages}{075125} (\bibinfo{year}{2013}),
  \urlprefix\url{https://link.aps.org/doi/10.1103/PhysRevB.88.075125}.

\bibitem[{\citenamefont{Hsieh et~al.}(2016)\citenamefont{Hsieh, Cho, and
  Ryu}}]{Hsieh16}
\bibinfo{author}{\bibfnamefont{C.-T.} \bibnamefont{Hsieh}},
  \bibinfo{author}{\bibfnamefont{G.~Y.} \bibnamefont{Cho}}, \bibnamefont{and}
  \bibinfo{author}{\bibfnamefont{S.}~\bibnamefont{Ryu}},
  \bibinfo{journal}{Phys. Rev. B} \textbf{\bibinfo{volume}{93}},
  \bibinfo{pages}{075135} (\bibinfo{year}{2016}),
  \urlprefix\url{https://link.aps.org/doi/10.1103/PhysRevB.93.075135}.

\bibitem[{\citenamefont{Sun et~al.}(2018)\citenamefont{Sun, Xiao, Bzdusek,
  Zhang, and Fan}}]{SunX18}
\bibinfo{author}{\bibfnamefont{X.-Q.} \bibnamefont{Sun}},
  \bibinfo{author}{\bibfnamefont{M.}~\bibnamefont{Xiao}},
  \bibinfo{author}{\bibfnamefont{T.}~\bibnamefont{Bzdusek}},
  \bibinfo{author}{\bibfnamefont{S.-C.} \bibnamefont{Zhang}}, \bibnamefont{and}
  \bibinfo{author}{\bibfnamefont{S.}~\bibnamefont{Fan}},
  \bibinfo{journal}{Phys. Rev. Lett.} \textbf{\bibinfo{volume}{121}},
  \bibinfo{pages}{196401} (\bibinfo{year}{2018}),
  \urlprefix\url{https://link.aps.org/doi/10.1103/PhysRevLett.121.196401}.

\bibitem[{\citenamefont{Budich et~al.}(2017)\citenamefont{Budich, Hu, and
  Zoller}}]{Budich17}
\bibinfo{author}{\bibfnamefont{J.~C.} \bibnamefont{Budich}},
  \bibinfo{author}{\bibfnamefont{Y.}~\bibnamefont{Hu}}, \bibnamefont{and}
  \bibinfo{author}{\bibfnamefont{P.}~\bibnamefont{Zoller}},
  \bibinfo{journal}{Phys. Rev. Lett.} \textbf{\bibinfo{volume}{118}},
  \bibinfo{pages}{105302} (\bibinfo{year}{2017}),
  \urlprefix\url{https://link.aps.org/doi/10.1103/PhysRevLett.118.105302}.

\bibitem[{\citenamefont{Nathan and Rudner}(2015)}]{NathanRudner15}
\bibinfo{author}{\bibfnamefont{F.}~\bibnamefont{Nathan}} \bibnamefont{and}
  \bibinfo{author}{\bibfnamefont{M.~S.} \bibnamefont{Rudner}},
  \bibinfo{journal}{New Journal of Physics} \textbf{\bibinfo{volume}{17}},
  \bibinfo{pages}{125014} (\bibinfo{year}{2015}),
  \urlprefix\url{http://stacks.iop.org/1367-2630/17/i=12/a=125014}.

\bibitem[{1d_()}]{1d_winding}
\bibinfo{note}{See Supplemental Material, appendix A for a detailed discussion
  on a single chiral fermion in a Floquet one-dimensional lattice.}

\bibitem[{3d_()}]{3d_winding}
\bibinfo{note}{See Supplemental Material, appendix B for the method of
  constructing a nontrivial map from $\mathbb{T}^3$ to $S^3$.}

\bibitem[{exp()}]{experiment}
\bibinfo{note}{See Supplemental Material, appendix C for experimental
  implementations by ultracold atoms.}

\bibitem[{\citenamefont{Manton and Sutcliffe}(2004)}]{MantonN04}
\bibinfo{author}{\bibfnamefont{N.}~\bibnamefont{Manton}} \bibnamefont{and}
  \bibinfo{author}{\bibfnamefont{P.}~\bibnamefont{Sutcliffe}},
  \emph{\bibinfo{title}{{Topological solitons}}} (\bibinfo{publisher}{Cambridge
  University Press}, \bibinfo{address}{Cambridge}, \bibinfo{year}{2004}).

\bibitem[{\citenamefont{Chan et~al.}(2016)\citenamefont{Chan, Oh, Han, and
  Lee}}]{Chan16a}
\bibinfo{author}{\bibfnamefont{C.-K.} \bibnamefont{Chan}},
  \bibinfo{author}{\bibfnamefont{Y.-T.} \bibnamefont{Oh}},
  \bibinfo{author}{\bibfnamefont{J.~H.} \bibnamefont{Han}}, \bibnamefont{and}
  \bibinfo{author}{\bibfnamefont{P.~A.} \bibnamefont{Lee}},
  \bibinfo{journal}{Phys. Rev. B} \textbf{\bibinfo{volume}{94}},
  \bibinfo{pages}{121106} (\bibinfo{year}{2016}),
  \urlprefix\url{https://link.aps.org/doi/10.1103/PhysRevB.94.121106}.

\bibitem[{\citenamefont{H{\"{u}}bener et~al.}(2017)\citenamefont{H{\"{u}}bener,
  Sentef, {De Giovannini}, Kemper, and Rubio}}]{Hubener17}
\bibinfo{author}{\bibfnamefont{H.}~\bibnamefont{H{\"{u}}bener}},
  \bibinfo{author}{\bibfnamefont{M.~A.} \bibnamefont{Sentef}},
  \bibinfo{author}{\bibfnamefont{U.}~\bibnamefont{{De Giovannini}}},
  \bibinfo{author}{\bibfnamefont{A.~F.} \bibnamefont{Kemper}},
  \bibnamefont{and} \bibinfo{author}{\bibfnamefont{A.}~\bibnamefont{Rubio}},
  \bibinfo{journal}{Nature Communications} \textbf{\bibinfo{volume}{8}},
  \bibinfo{pages}{13940} (\bibinfo{year}{2017}),
  \urlprefix\url{http://dx.doi.org/10.1038/ncomms13940
  http://10.0.4.14/ncomms13940
  https://www.nature.com/articles/ncomms13940{\#}supplementary-information}.

\bibitem[{\citenamefont{Wang et~al.}(2016{\natexlab{b}})\citenamefont{Wang,
  Zhou, and Chong}}]{WangH16}
\bibinfo{author}{\bibfnamefont{H.}~\bibnamefont{Wang}},
  \bibinfo{author}{\bibfnamefont{L.}~\bibnamefont{Zhou}}, \bibnamefont{and}
  \bibinfo{author}{\bibfnamefont{Y.~D.} \bibnamefont{Chong}},
  \bibinfo{journal}{Phys. Rev. B} \textbf{\bibinfo{volume}{93}},
  \bibinfo{pages}{144114} (\bibinfo{year}{2016}{\natexlab{b}}),
  \urlprefix\url{https://link.aps.org/doi/10.1103/PhysRevB.93.144114}.

\bibitem[{\citenamefont{Zou and Liu}(2016)}]{Zou16}
\bibinfo{author}{\bibfnamefont{J.-Y.} \bibnamefont{Zou}} \bibnamefont{and}
  \bibinfo{author}{\bibfnamefont{B.-G.} \bibnamefont{Liu}},
  \bibinfo{journal}{Phys. Rev. B} \textbf{\bibinfo{volume}{93}},
  \bibinfo{pages}{205435} (\bibinfo{year}{2016}),
  \urlprefix\url{https://link.aps.org/doi/10.1103/PhysRevB.93.205435}.

\bibitem[{\citenamefont{Bucciantini et~al.}(2017)\citenamefont{Bucciantini,
  Roy, Kitamura, and Oka}}]{Bucciantini17}
\bibinfo{author}{\bibfnamefont{L.}~\bibnamefont{Bucciantini}},
  \bibinfo{author}{\bibfnamefont{S.}~\bibnamefont{Roy}},
  \bibinfo{author}{\bibfnamefont{S.}~\bibnamefont{Kitamura}}, \bibnamefont{and}
  \bibinfo{author}{\bibfnamefont{T.}~\bibnamefont{Oka}},
  \bibinfo{journal}{Phys. Rev. B} \textbf{\bibinfo{volume}{96}},
  \bibinfo{pages}{041126} (\bibinfo{year}{2017}),
  \urlprefix\url{https://link.aps.org/doi/10.1103/PhysRevB.96.041126}.

\bibitem[{\citenamefont{Wang et~al.}(2014)\citenamefont{Wang, Wang, Shen,
  Sheng, and Xing}}]{Wang14}
\bibinfo{author}{\bibfnamefont{R.}~\bibnamefont{Wang}},
  \bibinfo{author}{\bibfnamefont{B.}~\bibnamefont{Wang}},
  \bibinfo{author}{\bibfnamefont{R.}~\bibnamefont{Shen}},
  \bibinfo{author}{\bibfnamefont{L.}~\bibnamefont{Sheng}}, \bibnamefont{and}
  \bibinfo{author}{\bibfnamefont{D.~Y.} \bibnamefont{Xing}},
  \bibinfo{journal}{EPL (Europhysics Letters)} \textbf{\bibinfo{volume}{105}},
  \bibinfo{pages}{17004} (\bibinfo{year}{2014}),
  \urlprefix\url{http://stacks.iop.org/0295-5075/105/i=1/a=17004}.

\bibitem[{\citenamefont{Ebihara et~al.}(2016)\citenamefont{Ebihara, Fukushima,
  and Oka}}]{Ebihara16}
\bibinfo{author}{\bibfnamefont{S.}~\bibnamefont{Ebihara}},
  \bibinfo{author}{\bibfnamefont{K.}~\bibnamefont{Fukushima}},
  \bibnamefont{and} \bibinfo{author}{\bibfnamefont{T.}~\bibnamefont{Oka}},
  \bibinfo{journal}{Phys. Rev. B} \textbf{\bibinfo{volume}{93}},
  \bibinfo{pages}{155107} (\bibinfo{year}{2016}),
  \urlprefix\url{https://link.aps.org/doi/10.1103/PhysRevB.93.155107}.

\bibitem[{\citenamefont{Zhang et~al.}(2016{\natexlab{b}})\citenamefont{Zhang,
  Ong, and Nagaosa}}]{ZhangNagaosa16}
\bibinfo{author}{\bibfnamefont{X.-X.} \bibnamefont{Zhang}},
  \bibinfo{author}{\bibfnamefont{T.~T.} \bibnamefont{Ong}}, \bibnamefont{and}
  \bibinfo{author}{\bibfnamefont{N.}~\bibnamefont{Nagaosa}},
  \bibinfo{journal}{Phys. Rev. B} \textbf{\bibinfo{volume}{94}},
  \bibinfo{pages}{235137} (\bibinfo{year}{2016}{\natexlab{b}}),
  \urlprefix\url{https://link.aps.org/doi/10.1103/PhysRevB.94.235137}.

\bibitem[{\citenamefont{Yan and Wang}(2016)}]{Yan16}
\bibinfo{author}{\bibfnamefont{Z.}~\bibnamefont{Yan}} \bibnamefont{and}
  \bibinfo{author}{\bibfnamefont{Z.}~\bibnamefont{Wang}},
  \bibinfo{journal}{Phys. Rev. Lett.} \textbf{\bibinfo{volume}{117}},
  \bibinfo{pages}{087402} (\bibinfo{year}{2016}),
  \urlprefix\url{https://link.aps.org/doi/10.1103/PhysRevLett.117.087402}.

\bibitem[{\citenamefont{Takasan et~al.}(2017)\citenamefont{Takasan, Nakagawa,
  and Kawakami}}]{Takasan17}
\bibinfo{author}{\bibfnamefont{K.}~\bibnamefont{Takasan}},
  \bibinfo{author}{\bibfnamefont{M.}~\bibnamefont{Nakagawa}}, \bibnamefont{and}
  \bibinfo{author}{\bibfnamefont{N.}~\bibnamefont{Kawakami}},
  \bibinfo{journal}{Phys. Rev. B} \textbf{\bibinfo{volume}{96}},
  \bibinfo{pages}{115120} (\bibinfo{year}{2017}),
  \urlprefix\url{https://link.aps.org/doi/10.1103/PhysRevB.96.115120}.

\bibitem[{\citenamefont{Sorensen et~al.}(2005)\citenamefont{Sorensen, Demler,
  and Lukin}}]{Sorensen05}
\bibinfo{author}{\bibfnamefont{A.~S.} \bibnamefont{Sorensen}},
  \bibinfo{author}{\bibfnamefont{E.}~\bibnamefont{Demler}}, \bibnamefont{and}
  \bibinfo{author}{\bibfnamefont{M.~D.} \bibnamefont{Lukin}},
  \bibinfo{journal}{Phys. Rev. Lett.} \textbf{\bibinfo{volume}{94}},
  \bibinfo{pages}{086803} (\bibinfo{year}{2005}),
  \urlprefix\url{https://link.aps.org/doi/10.1103/PhysRevLett.94.086803}.

\bibitem[{\citenamefont{Hafezi et~al.}(2007)\citenamefont{Hafezi, Sorensen,
  Demler, and Lukin}}]{HafeziM07}
\bibinfo{author}{\bibfnamefont{M.}~\bibnamefont{Hafezi}},
  \bibinfo{author}{\bibfnamefont{A.~S.} \bibnamefont{Sorensen}},
  \bibinfo{author}{\bibfnamefont{E.}~\bibnamefont{Demler}}, \bibnamefont{and}
  \bibinfo{author}{\bibfnamefont{M.~D.} \bibnamefont{Lukin}},
  \bibinfo{journal}{Phys. Rev. A} \textbf{\bibinfo{volume}{76}},
  \bibinfo{pages}{023613} (\bibinfo{year}{2007}),
  \urlprefix\url{https://link.aps.org/doi/10.1103/PhysRevA.76.023613}.

\bibitem[{bar()}]{barU_form}
\bibinfo{note}{See Supplemental Material, appendix D for the explicit form of
  $\bar{U}(k_2, k_3)$.}

\bibitem[{\citenamefont{Dauphin and Goldman}(2013)}]{DauphinA13}
\bibinfo{author}{\bibfnamefont{A.}~\bibnamefont{Dauphin}} \bibnamefont{and}
  \bibinfo{author}{\bibfnamefont{N.}~\bibnamefont{Goldman}},
  \bibinfo{journal}{Phys. Rev. Lett.} \textbf{\bibinfo{volume}{111}},
  \bibinfo{pages}{135302} (\bibinfo{year}{2013}),
  \urlprefix\url{https://link.aps.org/doi/10.1103/PhysRevLett.111.135302}.

\bibitem[{\citenamefont{Roy et~al.}(2016)\citenamefont{Roy, Kolodrubetz, Moore,
  and Grushin}}]{RoyS16}
\bibinfo{author}{\bibfnamefont{S.}~\bibnamefont{Roy}},
  \bibinfo{author}{\bibfnamefont{M.}~\bibnamefont{Kolodrubetz}},
  \bibinfo{author}{\bibfnamefont{J.~E.} \bibnamefont{Moore}}, \bibnamefont{and}
  \bibinfo{author}{\bibfnamefont{A.~G.} \bibnamefont{Grushin}},
  \bibinfo{journal}{Phys. Rev. B} \textbf{\bibinfo{volume}{94}},
  \bibinfo{pages}{161107} (\bibinfo{year}{2016}),
  \urlprefix\url{https://link.aps.org/doi/10.1103/PhysRevB.94.161107}.

\bibitem[{\citenamefont{Roy et~al.}(2018)\citenamefont{Roy, Kolodrubetz,
  Goldman, and Grushin}}]{RoyS18}
\bibinfo{author}{\bibfnamefont{S.}~\bibnamefont{Roy}},
  \bibinfo{author}{\bibfnamefont{M.}~\bibnamefont{Kolodrubetz}},
  \bibinfo{author}{\bibfnamefont{N.}~\bibnamefont{Goldman}}, \bibnamefont{and}
  \bibinfo{author}{\bibfnamefont{A.~G.} \bibnamefont{Grushin}},
  \bibinfo{journal}{2D Materials} \textbf{\bibinfo{volume}{5}},
  \bibinfo{pages}{24001} (\bibinfo{year}{2018}),
  \urlprefix\url{https://doi.org/10.1088{\%}2F2053-1583{\%}2Faaa577}.

\bibitem[{sin()}]{single_particle}
\bibinfo{note}{See Supplemental Material, appendix E for a single-particle
  dynamics under the Floquet operator.}

\bibitem[{\citenamefont{Kitaev}(2009)}]{Kitaev09}
\bibinfo{author}{\bibfnamefont{A.}~\bibnamefont{Kitaev}}, \bibinfo{journal}{AIP
  Conference Proceedings} \textbf{\bibinfo{volume}{1134}}, \bibinfo{pages}{22}
  (\bibinfo{year}{2009}),
  \eprint{http://aip.scitation.org/doi/pdf/10.1063/1.3149495},
  \urlprefix\url{http://aip.scitation.org/doi/abs/10.1063/1.3149495}.

\bibitem[{\citenamefont{Schnyder et~al.}(2008)\citenamefont{Schnyder, Ryu,
  Furusaki, and Ludwig}}]{Schnyder08}
\bibinfo{author}{\bibfnamefont{A.~P.} \bibnamefont{Schnyder}},
  \bibinfo{author}{\bibfnamefont{S.}~\bibnamefont{Ryu}},
  \bibinfo{author}{\bibfnamefont{A.}~\bibnamefont{Furusaki}}, \bibnamefont{and}
  \bibinfo{author}{\bibfnamefont{A.~W.~W.} \bibnamefont{Ludwig}},
  \bibinfo{journal}{Phys. Rev. B} \textbf{\bibinfo{volume}{78}},
  \bibinfo{pages}{195125} (\bibinfo{year}{2008}),
  \urlprefix\url{https://link.aps.org/doi/10.1103/PhysRevB.78.195125}.

\bibitem[{\citenamefont{Teo and Kane}(2010)}]{TeoKane10}
\bibinfo{author}{\bibfnamefont{J.~C.~Y.} \bibnamefont{Teo}} \bibnamefont{and}
  \bibinfo{author}{\bibfnamefont{C.~L.} \bibnamefont{Kane}},
  \bibinfo{journal}{Phys. Rev. B} \textbf{\bibinfo{volume}{82}},
  \bibinfo{pages}{115120} (\bibinfo{year}{2010}),
  \urlprefix\url{https://link.aps.org/doi/10.1103/PhysRevB.82.115120}.

\bibitem[{\citenamefont{Roy and Harper}(2017)}]{RoyHarper17}
\bibinfo{author}{\bibfnamefont{R.}~\bibnamefont{Roy}} \bibnamefont{and}
  \bibinfo{author}{\bibfnamefont{F.}~\bibnamefont{Harper}},
  \bibinfo{journal}{Phys. Rev. B} \textbf{\bibinfo{volume}{96}},
  \bibinfo{pages}{155118} (\bibinfo{year}{2017}),
  \urlprefix\url{https://link.aps.org/doi/10.1103/PhysRevB.96.155118}.

\bibitem[{\citenamefont{Morimoto et~al.}(2017)\citenamefont{Morimoto, Po, and
  Vishwanath}}]{MorimotoPoVishwanath17}
\bibinfo{author}{\bibfnamefont{T.}~\bibnamefont{Morimoto}},
  \bibinfo{author}{\bibfnamefont{H.~C.} \bibnamefont{Po}}, \bibnamefont{and}
  \bibinfo{author}{\bibfnamefont{A.}~\bibnamefont{Vishwanath}},
  \bibinfo{journal}{Phys. Rev. B} \textbf{\bibinfo{volume}{95}},
  \bibinfo{pages}{195155} (\bibinfo{year}{2017}),
  \urlprefix\url{https://link.aps.org/doi/10.1103/PhysRevB.95.195155}.

\bibitem[{der()}]{derivation_classification}
\bibinfo{note}{See Supplemental Material, appendix F for the detailed
  derivation of the classification in Table I.}

\bibitem[{\citenamefont{Qi et~al.}(2008)\citenamefont{Qi, Hughes, and
  Zhang}}]{Qi08}
\bibinfo{author}{\bibfnamefont{X.-L.} \bibnamefont{Qi}},
  \bibinfo{author}{\bibfnamefont{T.~L.} \bibnamefont{Hughes}},
  \bibnamefont{and} \bibinfo{author}{\bibfnamefont{S.-C.} \bibnamefont{Zhang}},
  \bibinfo{journal}{Phys. Rev. B} \textbf{\bibinfo{volume}{78}},
  \bibinfo{pages}{195424} (\bibinfo{year}{2008}),
  \urlprefix\url{https://link.aps.org/doi/10.1103/PhysRevB.78.195424}.

\bibitem[{\citenamefont{Nakahara}(2003)}]{Nakahara03}
\bibinfo{author}{\bibfnamefont{M.}~\bibnamefont{Nakahara}},
  \emph{\bibinfo{title}{{Geometry, topology and physics}}}
  (\bibinfo{publisher}{CRC Press}, \bibinfo{year}{2003}).

\bibitem[{\citenamefont{Mandel et~al.}(2003{\natexlab{a}})\citenamefont{Mandel,
  Greiner, Widera, Rom, H{\"{a}}nsch, and Bloch}}]{Mandel03a}
\bibinfo{author}{\bibfnamefont{O.}~\bibnamefont{Mandel}},
  \bibinfo{author}{\bibfnamefont{M.}~\bibnamefont{Greiner}},
  \bibinfo{author}{\bibfnamefont{A.}~\bibnamefont{Widera}},
  \bibinfo{author}{\bibfnamefont{T.}~\bibnamefont{Rom}},
  \bibinfo{author}{\bibfnamefont{T.~W.} \bibnamefont{H{\"{a}}nsch}},
  \bibnamefont{and} \bibinfo{author}{\bibfnamefont{I.}~\bibnamefont{Bloch}},
  \bibinfo{journal}{Nature} \textbf{\bibinfo{volume}{425}},
  \bibinfo{pages}{937} (\bibinfo{year}{2003}{\natexlab{a}}),
  \urlprefix\url{http://dx.doi.org/10.1038/nature02008
  http://10.0.4.14/nature02008}.

\bibitem[{\citenamefont{Mandel et~al.}(2003{\natexlab{b}})\citenamefont{Mandel,
  Greiner, Widera, Rom, H{\"{a}}nsch, and Bloch}}]{Mandel03b}
\bibinfo{author}{\bibfnamefont{O.}~\bibnamefont{Mandel}},
  \bibinfo{author}{\bibfnamefont{M.}~\bibnamefont{Greiner}},
  \bibinfo{author}{\bibfnamefont{A.}~\bibnamefont{Widera}},
  \bibinfo{author}{\bibfnamefont{T.}~\bibnamefont{Rom}},
  \bibinfo{author}{\bibfnamefont{T.~W.} \bibnamefont{H{\"{a}}nsch}},
  \bibnamefont{and} \bibinfo{author}{\bibfnamefont{I.}~\bibnamefont{Bloch}},
  \bibinfo{journal}{Phys. Rev. Lett.} \textbf{\bibinfo{volume}{91}},
  \bibinfo{pages}{010407} (\bibinfo{year}{2003}{\natexlab{b}}),
  \urlprefix\url{https://link.aps.org/doi/10.1103/PhysRevLett.91.010407}.

\bibitem[{\citenamefont{Jaksch and Zoller}(2003)}]{JakschD03}
\bibinfo{author}{\bibfnamefont{D.}~\bibnamefont{Jaksch}} \bibnamefont{and}
  \bibinfo{author}{\bibfnamefont{P.}~\bibnamefont{Zoller}},
  \bibinfo{journal}{New Journal of Physics} \textbf{\bibinfo{volume}{5}},
  \bibinfo{pages}{56} (\bibinfo{year}{2003}),
  \urlprefix\url{http://stacks.iop.org/1367-2630/5/i=1/a=356}.

\bibitem[{\citenamefont{Aidelsburger et~al.}(2014)\citenamefont{Aidelsburger,
  Lohse, Schweizer, Atala, Barreiro, Nascimb{\`{e}}ne, Cooper, Bloch, and
  Goldman}}]{AidelsburgerM14}
\bibinfo{author}{\bibfnamefont{M.}~\bibnamefont{Aidelsburger}},
  \bibinfo{author}{\bibfnamefont{M.}~\bibnamefont{Lohse}},
  \bibinfo{author}{\bibfnamefont{C.}~\bibnamefont{Schweizer}},
  \bibinfo{author}{\bibfnamefont{M.}~\bibnamefont{Atala}},
  \bibinfo{author}{\bibfnamefont{J.~T.} \bibnamefont{Barreiro}},
  \bibinfo{author}{\bibfnamefont{S.}~\bibnamefont{Nascimb{\`{e}}ne}},
  \bibinfo{author}{\bibfnamefont{N.~R.} \bibnamefont{Cooper}},
  \bibinfo{author}{\bibfnamefont{I.}~\bibnamefont{Bloch}}, \bibnamefont{and}
  \bibinfo{author}{\bibfnamefont{N.}~\bibnamefont{Goldman}},
  \bibinfo{journal}{Nature Physics} \textbf{\bibinfo{volume}{11}},
  \bibinfo{pages}{162} (\bibinfo{year}{2014}),
  \urlprefix\url{http://dx.doi.org/10.1038/nphys3171 http://10.0.4.14/nphys3171
  https://www.nature.com/articles/nphys3171{\#}supplementary-information}.

\bibitem[{\citenamefont{Goldman et~al.}(2014)\citenamefont{Goldman, Juzeliunas,
  {\"{O}}hberg, and Spielman}}]{Goldman14b}
\bibinfo{author}{\bibfnamefont{N.}~\bibnamefont{Goldman}},
  \bibinfo{author}{\bibfnamefont{G.}~\bibnamefont{Juzeliunas}},
  \bibinfo{author}{\bibfnamefont{P.}~\bibnamefont{{\"{O}}hberg}},
  \bibnamefont{and} \bibinfo{author}{\bibfnamefont{I.~B.}
  \bibnamefont{Spielman}}, \bibinfo{journal}{Reports on Progress in Physics}
  \textbf{\bibinfo{volume}{77}}, \bibinfo{pages}{126401}
  (\bibinfo{year}{2014}),
  \urlprefix\url{http://stacks.iop.org/0034-4885/77/i=12/a=126401}.

\bibitem[{\citenamefont{Bloch et~al.}(2008)\citenamefont{Bloch, Dalibard, and
  Zwerger}}]{Bloch08}
\bibinfo{author}{\bibfnamefont{I.}~\bibnamefont{Bloch}},
  \bibinfo{author}{\bibfnamefont{J.}~\bibnamefont{Dalibard}}, \bibnamefont{and}
  \bibinfo{author}{\bibfnamefont{W.}~\bibnamefont{Zwerger}},
  \bibinfo{journal}{Rev. Mod. Phys.} \textbf{\bibinfo{volume}{80}},
  \bibinfo{pages}{885} (\bibinfo{year}{2008}),
  \urlprefix\url{https://link.aps.org/doi/10.1103/RevModPhys.80.885}.

\bibitem[{\citenamefont{Mancini et~al.}(2015)\citenamefont{Mancini, Pagano,
  Cappellini, Livi, Rider, Catani, Sias, Zoller, Inguscio, Dalmonte
  et~al.}}]{Mancini15}
\bibinfo{author}{\bibfnamefont{M.}~\bibnamefont{Mancini}},
  \bibinfo{author}{\bibfnamefont{G.}~\bibnamefont{Pagano}},
  \bibinfo{author}{\bibfnamefont{G.}~\bibnamefont{Cappellini}},
  \bibinfo{author}{\bibfnamefont{L.}~\bibnamefont{Livi}},
  \bibinfo{author}{\bibfnamefont{M.}~\bibnamefont{Rider}},
  \bibinfo{author}{\bibfnamefont{J.}~\bibnamefont{Catani}},
  \bibinfo{author}{\bibfnamefont{C.}~\bibnamefont{Sias}},
  \bibinfo{author}{\bibfnamefont{P.}~\bibnamefont{Zoller}},
  \bibinfo{author}{\bibfnamefont{M.}~\bibnamefont{Inguscio}},
  \bibinfo{author}{\bibfnamefont{M.}~\bibnamefont{Dalmonte}},
  \bibnamefont{et~al.}, \bibinfo{journal}{Science}
  \textbf{\bibinfo{volume}{349}}, \bibinfo{pages}{1510} (\bibinfo{year}{2015}),
  ISSN \bibinfo{issn}{10959203}, \eprint{1502.02495},
  \urlprefix\url{http://arxiv.org/abs/1502.02495}.

\bibitem[{\citenamefont{Song et~al.}(2016)\citenamefont{Song, He, Zhang,
  Hajiyev, Huang, Liu, and Jo}}]{Song16}
\bibinfo{author}{\bibfnamefont{B.}~\bibnamefont{Song}},
  \bibinfo{author}{\bibfnamefont{C.}~\bibnamefont{He}},
  \bibinfo{author}{\bibfnamefont{S.}~\bibnamefont{Zhang}},
  \bibinfo{author}{\bibfnamefont{E.}~\bibnamefont{Hajiyev}},
  \bibinfo{author}{\bibfnamefont{W.}~\bibnamefont{Huang}},
  \bibinfo{author}{\bibfnamefont{X.-J.} \bibnamefont{Liu}}, \bibnamefont{and}
  \bibinfo{author}{\bibfnamefont{G.-B.} \bibnamefont{Jo}},
  \bibinfo{journal}{Phys. Rev. A} \textbf{\bibinfo{volume}{94}},
  \bibinfo{pages}{061604} (\bibinfo{year}{2016}),
  \urlprefix\url{https://link.aps.org/doi/10.1103/PhysRevA.94.061604}.

\bibitem[{\citenamefont{Budich et~al.}(2015)\citenamefont{Budich, Laflamme,
  Tschirsich, Montangero, and Zoller}}]{BudichJ15}
\bibinfo{author}{\bibfnamefont{J.~C.} \bibnamefont{Budich}},
  \bibinfo{author}{\bibfnamefont{C.}~\bibnamefont{Laflamme}},
  \bibinfo{author}{\bibfnamefont{F.}~\bibnamefont{Tschirsich}},
  \bibinfo{author}{\bibfnamefont{S.}~\bibnamefont{Montangero}},
  \bibnamefont{and} \bibinfo{author}{\bibfnamefont{P.}~\bibnamefont{Zoller}},
  \bibinfo{journal}{Phys. Rev. B} \textbf{\bibinfo{volume}{92}},
  \bibinfo{pages}{245121} (\bibinfo{year}{2015}),
  \urlprefix\url{https://link.aps.org/doi/10.1103/PhysRevB.92.245121}.

\bibitem[{\citenamefont{Enomoto et~al.}(2008)\citenamefont{Enomoto, Kasa,
  Kitagawa, and Takahashi}}]{Enomoto08}
\bibinfo{author}{\bibfnamefont{K.}~\bibnamefont{Enomoto}},
  \bibinfo{author}{\bibfnamefont{K.}~\bibnamefont{Kasa}},
  \bibinfo{author}{\bibfnamefont{M.}~\bibnamefont{Kitagawa}}, \bibnamefont{and}
  \bibinfo{author}{\bibfnamefont{Y.}~\bibnamefont{Takahashi}},
  \bibinfo{journal}{Phys. Rev. Lett.} \textbf{\bibinfo{volume}{101}},
  \bibinfo{pages}{203201} (\bibinfo{year}{2008}),
  \urlprefix\url{https://link.aps.org/doi/10.1103/PhysRevLett.101.203201}.

\bibitem[{\citenamefont{Aubry and Andr{\'{e}}}(1980)}]{Aubry80}
\bibinfo{author}{\bibfnamefont{S.}~\bibnamefont{Aubry}} \bibnamefont{and}
  \bibinfo{author}{\bibfnamefont{G.}~\bibnamefont{Andr{\'{e}}}},
  \bibinfo{journal}{Ann. Israel Phys. Soc} \textbf{\bibinfo{volume}{3}},
  \bibinfo{pages}{18} (\bibinfo{year}{1980}).

\bibitem[{\citenamefont{Harper}(1955)}]{Harper55}
\bibinfo{author}{\bibfnamefont{P.~G.} \bibnamefont{Harper}},
  \bibinfo{journal}{Proceedings of the Physical Society. Section A}
  \textbf{\bibinfo{volume}{68}}, \bibinfo{pages}{874} (\bibinfo{year}{1955}),
  \urlprefix\url{http://stacks.iop.org/0370-1298/68/i=10/a=304}.

\bibitem[{\citenamefont{Gao et~al.}(2016)\citenamefont{Gao, Gao, Shi, Yang,
  Lin, Xu, Joannopoulos, Solja{\v{c}}i{\'{c}}, Chen, Lu et~al.}}]{GaoF16}
\bibinfo{author}{\bibfnamefont{F.}~\bibnamefont{Gao}},
  \bibinfo{author}{\bibfnamefont{Z.}~\bibnamefont{Gao}},
  \bibinfo{author}{\bibfnamefont{X.}~\bibnamefont{Shi}},
  \bibinfo{author}{\bibfnamefont{Z.}~\bibnamefont{Yang}},
  \bibinfo{author}{\bibfnamefont{X.}~\bibnamefont{Lin}},
  \bibinfo{author}{\bibfnamefont{H.}~\bibnamefont{Xu}},
  \bibinfo{author}{\bibfnamefont{J.~D.} \bibnamefont{Joannopoulos}},
  \bibinfo{author}{\bibfnamefont{M.}~\bibnamefont{Solja{\v{c}}i{\'{c}}}},
  \bibinfo{author}{\bibfnamefont{H.}~\bibnamefont{Chen}},
  \bibinfo{author}{\bibfnamefont{L.}~\bibnamefont{Lu}}, \bibnamefont{et~al.},
  \bibinfo{journal}{Nature Communications} \textbf{\bibinfo{volume}{7}},
  \bibinfo{pages}{11619} (\bibinfo{year}{2016}),
  \urlprefix\url{http://dx.doi.org/10.1038/ncomms11619
  http://10.0.4.14/ncomms11619
  https://www.nature.com/articles/ncomms11619{\#}supplementary-information}.

\bibitem[{\citenamefont{Mukherjee et~al.}(2017)\citenamefont{Mukherjee,
  Spracklen, Valiente, Andersson, {\"{O}}hberg, Goldman, and
  Thomson}}]{MukherjeeS17}
\bibinfo{author}{\bibfnamefont{S.}~\bibnamefont{Mukherjee}},
  \bibinfo{author}{\bibfnamefont{A.}~\bibnamefont{Spracklen}},
  \bibinfo{author}{\bibfnamefont{M.}~\bibnamefont{Valiente}},
  \bibinfo{author}{\bibfnamefont{E.}~\bibnamefont{Andersson}},
  \bibinfo{author}{\bibfnamefont{P.}~\bibnamefont{{\"{O}}hberg}},
  \bibinfo{author}{\bibfnamefont{N.}~\bibnamefont{Goldman}}, \bibnamefont{and}
  \bibinfo{author}{\bibfnamefont{R.~R.} \bibnamefont{Thomson}},
  \bibinfo{journal}{Nature Communications} \textbf{\bibinfo{volume}{8}},
  \bibinfo{pages}{13918} (\bibinfo{year}{2017}),
  \urlprefix\url{http://dx.doi.org/10.1038/ncomms13918
  http://10.0.4.14/ncomms13918
  https://www.nature.com/articles/ncomms13918{\#}supplementary-information}.

\bibitem[{\citenamefont{Maczewsky et~al.}(2017)\citenamefont{Maczewsky, Zeuner,
  Nolte, and Szameit}}]{MaczewskyL17}
\bibinfo{author}{\bibfnamefont{L.~J.} \bibnamefont{Maczewsky}},
  \bibinfo{author}{\bibfnamefont{J.~M.} \bibnamefont{Zeuner}},
  \bibinfo{author}{\bibfnamefont{S.}~\bibnamefont{Nolte}}, \bibnamefont{and}
  \bibinfo{author}{\bibfnamefont{A.}~\bibnamefont{Szameit}},
  \bibinfo{journal}{Nature Communications} \textbf{\bibinfo{volume}{8}},
  \bibinfo{pages}{13756} (\bibinfo{year}{2017}),
  \urlprefix\url{https://doi.org/10.1038/ncomms13756
  http://10.0.4.14/ncomms13756
  https://www.nature.com/articles/ncomms13756{\#}supplementary-information}.

\bibitem[{\citenamefont{Kitagawa
  et~al.}(2010{\natexlab{b}})\citenamefont{Kitagawa, Rudner, Berg, and
  Demler}}]{KitagawaT10}
\bibinfo{author}{\bibfnamefont{T.}~\bibnamefont{Kitagawa}},
  \bibinfo{author}{\bibfnamefont{M.~S.} \bibnamefont{Rudner}},
  \bibinfo{author}{\bibfnamefont{E.}~\bibnamefont{Berg}}, \bibnamefont{and}
  \bibinfo{author}{\bibfnamefont{E.}~\bibnamefont{Demler}},
  \bibinfo{journal}{Phys. Rev. A} \textbf{\bibinfo{volume}{82}},
  \bibinfo{pages}{033429} (\bibinfo{year}{2010}{\natexlab{b}}),
  \urlprefix\url{https://link.aps.org/doi/10.1103/PhysRevA.82.033429}.

\bibitem[{\citenamefont{Kitagawa et~al.}(2012)\citenamefont{Kitagawa, Broome,
  Fedrizzi, Rudner, Berg, Kassal, Aspuru-Guzik, Demler, and
  White}}]{KitagawaT12}
\bibinfo{author}{\bibfnamefont{T.}~\bibnamefont{Kitagawa}},
  \bibinfo{author}{\bibfnamefont{M.~A.} \bibnamefont{Broome}},
  \bibinfo{author}{\bibfnamefont{A.}~\bibnamefont{Fedrizzi}},
  \bibinfo{author}{\bibfnamefont{M.~S.} \bibnamefont{Rudner}},
  \bibinfo{author}{\bibfnamefont{E.}~\bibnamefont{Berg}},
  \bibinfo{author}{\bibfnamefont{I.}~\bibnamefont{Kassal}},
  \bibinfo{author}{\bibfnamefont{A.}~\bibnamefont{Aspuru-Guzik}},
  \bibinfo{author}{\bibfnamefont{E.}~\bibnamefont{Demler}}, \bibnamefont{and}
  \bibinfo{author}{\bibfnamefont{A.~G.} \bibnamefont{White}},
  \bibinfo{journal}{Nature Communications} \textbf{\bibinfo{volume}{3}},
  \bibinfo{pages}{882} (\bibinfo{year}{2012}),
  \urlprefix\url{http://dx.doi.org/10.1038/ncomms1872
  http://10.0.4.14/ncomms1872}.

\bibitem[{\citenamefont{Jotzu et~al.}(2014)\citenamefont{Jotzu, Messer,
  Desbuquois, Lebrat, Uehlinger, Greif, and Esslinger}}]{JotzuG14}
\bibinfo{author}{\bibfnamefont{G.}~\bibnamefont{Jotzu}},
  \bibinfo{author}{\bibfnamefont{M.}~\bibnamefont{Messer}},
  \bibinfo{author}{\bibfnamefont{R.}~\bibnamefont{Desbuquois}},
  \bibinfo{author}{\bibfnamefont{M.}~\bibnamefont{Lebrat}},
  \bibinfo{author}{\bibfnamefont{T.}~\bibnamefont{Uehlinger}},
  \bibinfo{author}{\bibfnamefont{D.}~\bibnamefont{Greif}}, \bibnamefont{and}
  \bibinfo{author}{\bibfnamefont{T.}~\bibnamefont{Esslinger}},
  \bibinfo{journal}{Nature} \textbf{\bibinfo{volume}{515}},
  \bibinfo{pages}{237} (\bibinfo{year}{2014}),
  \urlprefix\url{http://dx.doi.org/10.1038/nature13915
  http://10.0.4.14/nature13915
  https://www.nature.com/articles/nature13915{\#}supplementary-information}.

\bibitem[{\citenamefont{Stuhl et~al.}(2015)\citenamefont{Stuhl, Lu, Aycock,
  Genkina, and Spielman}}]{Stuhl15}
\bibinfo{author}{\bibfnamefont{B.~K.} \bibnamefont{Stuhl}},
  \bibinfo{author}{\bibfnamefont{H.-I.} \bibnamefont{Lu}},
  \bibinfo{author}{\bibfnamefont{L.~M.} \bibnamefont{Aycock}},
  \bibinfo{author}{\bibfnamefont{D.}~\bibnamefont{Genkina}}, \bibnamefont{and}
  \bibinfo{author}{\bibfnamefont{I.~B.} \bibnamefont{Spielman}},
  \bibinfo{journal}{Science} \textbf{\bibinfo{volume}{349}},
  \bibinfo{pages}{1514} (\bibinfo{year}{2015}), ISSN \bibinfo{issn}{0036-8075},
  \urlprefix\url{http://science.sciencemag.org/content/349/6255/1514}.

\bibitem[{\citenamefont{Schr{\"{o}}dinger}(1930)}]{SchrodingerE30}
\bibinfo{author}{\bibfnamefont{E.}~\bibnamefont{Schr{\"{o}}dinger}},
  \bibinfo{journal}{Sitz. Preuss. Akad. Wiss. Phys.—Math. Kl.}
  \textbf{\bibinfo{volume}{24}}, \bibinfo{pages}{418} (\bibinfo{year}{1930}).

\bibitem[{\citenamefont{Bermudez et~al.}(2007)\citenamefont{Bermudez,
  Martin-Delgado, and Solano}}]{BermudezA07}
\bibinfo{author}{\bibfnamefont{A.}~\bibnamefont{Bermudez}},
  \bibinfo{author}{\bibfnamefont{M.~A.} \bibnamefont{Martin-Delgado}},
  \bibnamefont{and} \bibinfo{author}{\bibfnamefont{E.}~\bibnamefont{Solano}},
  \bibinfo{journal}{Phys. Rev. Lett.} \textbf{\bibinfo{volume}{99}},
  \bibinfo{pages}{123602} (\bibinfo{year}{2007}),
  \urlprefix\url{https://link.aps.org/doi/10.1103/PhysRevLett.99.123602}.

\bibitem[{\citenamefont{Rusin and Zawadzki}(2010)}]{RusinT10}
\bibinfo{author}{\bibfnamefont{T.~M.} \bibnamefont{Rusin}} \bibnamefont{and}
  \bibinfo{author}{\bibfnamefont{W.}~\bibnamefont{Zawadzki}},
  \bibinfo{journal}{Phys. Rev. D} \textbf{\bibinfo{volume}{82}},
  \bibinfo{pages}{125031} (\bibinfo{year}{2010}),
  \urlprefix\url{https://link.aps.org/doi/10.1103/PhysRevD.82.125031}.

\end{thebibliography}
\end{document}